\documentclass[twocolumn]{aastex63}
\submitjournal{AJ}

\shortauthors{Vokrouhlick{\' y} et~al.}

\usepackage{amsmath}

\graphicspath{{./}{figures/}}

\usepackage{tikz}
\usetikzlibrary{arrows,shapes,positioning,shadows,trees}

\tikzset{
  basic/.style  = {draw, text width=4cm, drop shadow, font=\sffamily, rectangle},
  root/.style   = {basic, rounded corners=2pt, thin, align=center, fill=blue!10},
  level 2/.style = {basic, rounded corners=6pt, thin,align=center, fill=pink!40, text width=11em},
  level 3/.style = {basic, thin, align=center, fill=white!60, text width=8em}
}

\usepackage[utf8]{inputenc}
\usepackage{dirtytalk}

\newcommand{\cC}{\mathcal{C}}

\begin{document}

\title{Analysis of Karin and Koronis2 asteroid families: new findings and challenges}

\correspondingauthor{David Vokrouhlick\'y}
\email{vokrouhl@cesnet.cz}

\author[0000-0002-6034-5452]{David Vokrouhlick\'y}
\affiliation{Astronomical Institute, Charles University, V Hole\v{s}ovi\v{c}k\'ach 2,
             CZ 18000, Prague 8, Czech Republic}
\author[0000-0002-4547-4301]{David Nesvorn{\' y}}
\affiliation{Department of Space Studies, Southwest Research Institute, 1301 Walnut St., Suite 400,
             Boulder, CO 80302, United States}
\author[0000-0002-1804-7814]{William F. Bottke}
\affiliation{Department of Space Studies, Southwest Research Institute, 1301 Walnut St., Suite 400,
             Boulder, CO 80302, United States}

\begin{abstract}
We use our home catalog of the asteroid proper elements to study the Karin family. The hierarchical clustering method provides formal identification with 3,863 members, but this set also includes objects from the neighboring Koronis2 and Kuitaisi families, as well as interlopers originating from the much older Koronis family. By tracking the trajectories of cluster objects backward in time, we identified 2,161 asteroids whose orbits converged with that of their parent body (832) Karin at $5.72\pm 0.09$~My ago ($95$\% C.L.). This method of calculating the family's age is based on a novel convergence metric that is directly related to the velocities at which fragments were ejected from (832)~Karin. We analyze the extent to which members $\leq 1.5$~km in diameter had drifted in semimajor axis due to Yarkovsky thermal forces and find it reflects the tilt of their rotation poles away from the ecliptic, recording the influence of the YORP torque. Karin's size frequency distribution in the $\simeq(0.8-3)$~km range follows a power-law with a cumulative slope index $-3.20\pm 0.01$. Removing members of the Karin family from the original group, we examine the Koronis2 family, whose members are associated with (158)~Koronis. We find it difficult for large members of the Koronis2 family to converge with the orbit of (158)~Koronis within its previously estimated age of $7.6$~My. Achieving such convergence would require the Koronis2 family to be older than $10$~My, but our result must be verified with a direct numerical approach in the future.
\end{abstract}


\section{Introduction} \label{intro}
The main asteroid belt contains several hundred clusters of collisional origin \citep[e.g.,][]{mil2014,netal2015,nov2022,nes2024}. For the most part, they have been identified in proper elements space, defined by semimajor axis $a$, eccentricity $e$ and sine of inclination $\sin I$, using the hierarchical clustering method (HCM) developed in the early 1990's \citep[e.g.,][]{zap1990,netal2015}. Although powerful, the HCM approach often combines fragments from a given  collisional event with nearby background asteroids or, as in the case of our work, with nearby families \citep[e.g.,][]{mil2014,nov2022}. In such cases, further analysis is required to distinguish and isolate the substructures from one another. Spectroscopic or albedo information may be useful in this respect \citep[e.g.,][]{metal2015,netal2015}, but if the background population or overlapping families share the same physical characteristics, other tools are needed. In certain situations, the age of a family can serve as a distinguishing factor. Within this context, the Karin family represents an archetype of this class \citep[with further examples such as the similarly young collisional cluster around the outer main belt active asteroid 300163 (2006~VW139) embodied in the Themis family; e.g.,][]{nov2012}.

The discovery of the Karin asteroid family by \citet{nes2002} marked the beginning of a new era in the century-long study of collisionally formed clusters within the main belt. For the first time, the epoch of the family's origin was determined through the direct backward propagation of its members' orbits. The success of this approach was the result of a fortunate combination of several factors unique to the Karin family: (i) its relatively young age of $\simeq 5.8$~My, (ii) its position within one of the most dynamically stable orbital zones in the asteroid main belt, and (iii) the relatively large size of its parent body, measuring $\simeq 33$~km in diameter, which produced a sufficient number of multikilometer-sized fragments. Since its discovery, the Karin family has been revisited numerous times \citep[e.g.,][]{nes2003,nes2004,nes2006,nes2006bands,car2016}, with each study uncovering new insights about this unique cluster. These findings have not only revealed previously unknown processes that can shape asteroid families but have also given us insights into whether young families are a plausible source for some meteorites.   
 
In this study, we use the catalog of asteroid proper orbital elements by \citet{nes2024} to investigate the Karin family.%
\footnote{Very similar results would have been obtained with the proper orbital elements available at the Asteroid Families Portal that currently contains as many inputs \citep[see][]{nov2022}.}
This effort is especially timely given recent discoveries showing that the Karin orbital zone contains several other collisional clusters, such as the Koronis2 family associated with (158) Koronis, that are of significant scientific interest \citep[see, e.g.,][]{molnar2009,nes2024,broz2024,nes2025}.

In the case of young families, defined here as those approximately younger than 10 My old, the initial orbital configuration may be reached by propagating the heliocentric orbits of individual members backward in time. The most computationally demanding approach involves the backward integration of heliocentric osculating orbits (or heliocentric positions and velocities). In Appendix~\ref{kamean}, we present a novel method for characterizing a family's origin using mean orbital elements and then apply it to the Karin family. This approach though requires detailed bakward numerical propagation of osculating orbits in the first place, and thus represents the same computer-time load.

As a first step, however, we find it more practical to employ a less complex method that reconstructs the past configuration of an asteroid family in a simpler orbital framework. In particular, we use the proper orbital element space $(a,e,\sin I)$, extended by the proper values of the longitude of node $\Omega$ and perihelion $\varpi$. Although this set cannot capture the full details of each asteroid's orbital history, the use of this space does allow us to efficiently reconstruct the family's past orbital configuration with minimal computational demands (in particular, without having to perform new N-body integrations).

Given that the proper elements $(a,e,\sin I)$ are already highly clustered and, by definition, remain constant over time (with the exception of semimajor axis $a$ as discussed below), the primary focus of reconstruction lies in the behavior of proper $\Omega$ and $\varpi$. While their current values may appear randomly distributed between $0^\circ$ and $360^\circ$, they must have been highly clustered at the time of the family's formation (see Sec.~\ref{ka3}). Achieving convergence of these angles in a past epoch not only determines the age of the family but also enables us to identify and remove background asteroids and/or members of overlapping families.

The structure of our paper is as follows.  In Sec.~\ref{theor}, we provide an overview of our method, which is fully applied to the state-of-the-art characterization of the Karin family in Sec.~\ref{res}. In Sec.~\ref{ka3}, we introduce a new approach to convergence analysis for young families, where the minimization function is directly linked to the velocity field with which the family members were initially dispersed after their parent body's disruption. We test this concept on the Karin family in Appendix~\ref{kamean}, though a comprehensive application of the method is left for future work. After isolating the true Karin family from the cluster identified by the HCM, we briefly examine the membership of the Koronis2 family within the remnant population (Sec.~\ref{koronis2}). While we highlight some intriguing puzzles associated with this important cluster (see also Sec.~\ref{sfdkk}), we defer its detailed analysis to a separate study. Our findings are summarized in the concluding Sec.~\ref{concl}.


\section{Methods}\label{theor}
The process of determining the age of a family using the past orbital convergence of its members can be summarized as follows \citep[for more details, see][]{nes2004,car2016}. Consider a set of $N+1$ asteroids that are candidate members of the family under study (either the Karin or Koronis2 family in this work). Each asteroid is assigned an index $j$, where $j=0$ represents the reference body (the largest remnant in the family), and $j=1,\dots,N$ correspond to the remaining asteroids, typically ordered by increasing absolute magnitude $H$ (or equivalently, decreasing size $D$).  

We outline the procedure using the proper longitude of the ascending node, $\Omega_j$, although the same process applies to the proper longitude of perihelion, $\varpi_j$. The current difference in proper longitude between each asteroid and the reference body is denoted as $\Delta \Omega_j(0)=\Omega_j(0)-\Omega_0(0)$. These values are randomly distributed throughout the definition interval. To project $\Delta\Omega_j$ back to a past epoch $T$ (with $T$ being positive in our notation), we utilize the associated proper frequencies, $s_j$ and $s_0$, which are provided alongside the proper elements. This process allows us to assess the degree of orbital convergence at earlier times. From that, we have $\Delta \Omega_j\left(T\right) =\Delta \Omega_j\left(0\right) + \left(s_0-s_j\right)\,T$ (the order of $s$ frequencies in the bracket of the second term is reversed because a positive time $T$ represents the past history). 

A complication for observed families, however, is that the orbits of kilometer-sized asteroids are perturbed by thermal accelerations known as the Yarkovsky effect \citep[e.g.][]{vetal2015}. The primary secular effect pertains to the semimajor axis 
$a$, for which we approximate a linear change over time as a first-order assumption i.e. $a_j(T)=a_j(0)-\dot{a}_j\,T$. The rate of change, $\dot{a}_j$, is unique to each asteroid and depends primarily on its size and the obliquity of its spin axis \citep[negative for retrograde-rotating asteroids and positive for prograde-rotating asteroids, see][]{vetal2015}. We neglect the small proper eccentricity and inclination changes caused by the Yarkovsky effect. Over timescales of $< 10$~Myr, the semimajor axis changes by a negligible amount, even for kilometer-sized objects in the Karin/Koronis2 region, and as such, $a_j(T)$ is not used as the primary metric for orbital convergence. Instead, the focus is on the indirect effect this change has on proper frequencies (such as $s_j$), which are highly sensitive to variations in $a_j$ and play a dominant role in the analysis \citep[e.g.,][]{nes2004}. The nodal difference $\Delta \Omega_j(T)$ must therefore account for the effect of changes in the semimajor axis and can be expressed as:
\begin{eqnarray}
 \Delta \Omega_j\left(T\right) &=& \Delta \Omega_j\left(0\right) + \left(s_0-s_j\right)\,T+\nonumber \\ & & 
 \frac{1}{2}\left[\left(\frac{\partial s}{\partial a}\right)_j\dot{a}_j-\left(\frac{\partial s}{\partial a}\right)_0\dot{a}_0\right]\,T^2\; .\label{conv1}
\end{eqnarray}

In principle, higher-order derivatives of $s$ with respect to $a$ should appear in the right hand side of Eq.~(\ref{conv1}), but their values are small and may be neglected for $T<10$~My. The first derivatives, however, must be known from analytical formulations or from estimates based on numerical analysis. We took the second option and found that a constant value $(\partial s/\partial a)=-69.0\pm 1.3$ arcsec~yr$^{-1}$ provides an adequate approximation for our purposes, with the uncertainty reflecting the range of values inferred in the Karin family region. Similarly, for the proper perihelion frequency $g$, we found that $(\partial g/\partial a)=85.6\pm 3.7$ arcsec~yr$^{-1}$ offers a satisfactory approximation, which we apply to all asteroids within the Karin orbital space.

Combining the above formulation for $\Omega_j$ with that of $\varpi_j$, we define a target function $\cC(T)$ to characterize the degree of convergence of the proper secular angles at any past epoch $T$ analyzed in the simulation.
\begin{equation} 
  \cC^2\left(T\right) = \cC^2_\Omega\left(T\right)+\cC^2_\varpi\left(T\right) = \sum_{j=1}^N \left[\Delta \Omega_j^2\left(T\right)+\Delta \varpi_j^2\left(T\right)\right] , \label{tf}
\end{equation}
where the summation includes all family fragments, beginning with the second-largest remnant. The parts $\cC_\Omega$ and $\cC_\varpi$ in $\cC$ denote individual contributions from the convergence of the nodes and perihelia. 

We also define averages ${\cC^\prime}^2=\cC^2/(2N)$, ${\cC_\Omega^\prime}^2=\cC_\Omega^2/N$ and ${\cC_\varpi^\prime}^2=\cC_\varpi^2/N$. The uniformly distributed angles $\Delta \Omega_j$ and $\Delta \varpi_j$ would have $\cC^\prime=180^\circ/\sqrt{3}\simeq 104^\circ$. In contrast, at the formation of the Karin or Koronis2 families, we expect $\cC^\prime\simeq 1^\circ-2^\circ$ (Sec.~\ref{ka3}). 

The goal of the initial family reconstruction method is to seek an epoch $T$ in the past such that $\cC^\prime$ in (\ref{tf}) with $\Delta\Omega_j(T)$ substituted from (\ref{conv1}), and similarly for perihelia differences, collapses to the expected degree-level. Some candidate asteroids from the originally analyzed sample may have secular angles that fail to converge at the epoch where the majority converge. These are interlopers that must be iteratively removed from the list. As with most optimization problems, the solution is achieved by adjusting a set of free parameters.  In our case, we consider the a priori unknown drift rates $\dot{a}_j$. Interestingly, their fitted values can be independently verified a posteriori through observations, particularly by analyzing photometric data to determine the rotation states of the asteroids.

\section{Results}\label{res}
We utilized our home catalog of asteroid proper orbital elements from \citet{nes2024} to identify the Karin family using the HCM with a velocity cutoff $10$ m~s$^{-1}$ cutoff velocity \citep[the order of magnitude by which the currently observed Karin members are separated from (832)~Karin; e.g.,][]{nes2006}. The resulting cluster comprises 3,863 objects, which include members of both the Karin and Koronis2 families, as well as potential interlopers from the older Koronis family and the younger Kuitaisi/Koronis3 families. Notably, the same cluster is obtained when starting the HCM method from (158)~Koronis and applying the same velocity cutoff. This result stems from the overlap between the two families. 

In the following analysis, we apply the methods outlined in Sec.~\ref{theor} to separate the Karin and Koronis2 families from the identified asteroid cluster. We first focus on identifying the Karin family, given that it is well-established, it has a younger age, and that it has well-constrained properties (Sec.~\ref{karin}). Once the Karin members are removed from this cluster, we will analyze the remaining asteroid sample to identify the members of the Koronis2 family (Sec.~\ref{koronis2}).

\subsection{Karin family}\label{karin}

\subsubsection{Step~1: Analysis of large member population}\label{ka1}
Since its discovery in 2002 \citep{nes2002}, the population of large members within the Karin family has been extensively studied. The main results relevant for our work are as follows:
\begin{itemize}
\item The physical parameters of the largest family member (832) Karin are well determined: (i) its size is $16.3$~km, based on an analysis of Wide-field Infrared Survey Explorer (WISE) observations \citep[e.g.,][]{mas2011}, (ii) its slow rotation period of $18.35$~hr and pole obliquity of $42^\circ$ were both constrained by analysis of photometric observations \citep[e.g.,][]{sli2012}. Using this information, and assuming it has a $2.5$ g~cm$^{-3}$ bulk density, which is typical for S-type asteroids of this size \cite[e.g.,][]{setal2015}, we estimate that its Yarkovsky drift rate is $\dot{a}_0\simeq 1.2\times 10^{-5}$ au~My$^{-1}$. This value will be used in all the simulations discussed in this paper.

\item The past orbital convergence of Karin's large members yielded a unique age solution for the Karin family, namely $T=5.75\pm 0.05$~My by \citet{nes2004} or $T=5.75\pm 0.01$~My by \citet{car2016}. The uncertainty of the latter should be taken as a formal one; a more realistic value would likely be an order of magnitude larger.
\item The hydrocode simulations of the impact event that formed the Karin family, conducted by \citet{nes2006} and calibrated based on their success in reproducing the size distribution of the largest members, predict that the mean ejection velocity for $D\geq 3$~km size fragments was about 10 m~s$^{-1}$. It reached values larger than $30$ m~s$^{-1}$ for kilometer-sized fragments.
\end{itemize}
To ensure a robust analysis, we began by focusing on a population of large asteroids among the $3,863$ objects identified as members of the Karin cluster by the HCM. Specifically, we selected all objects with an absolute magnitude of $H\leq 16.7$, corresponding to a size of approximately 1.3~km assuming a geometric albedo of 0.23. This initial selection resulted in 646 asteroids.

To refine the sample further, we excluded objects located significantly beyond the well-defined boundaries of the Karin family in two-dimensional planes defined by (i) $a$ and proper $e$ orbital elements, and (ii) proper $a$ and absolute magnitude $H$. After this step, we narrowed the sample down to $421$ candidate members of the Karin family, including the asteroid (832)~Karin, within the chosen $H$ limit.

Next, we applied the method described in Sec.~\ref{theor} to search for convergence of the proper nodes $\Omega$ and perihelia $\varpi$ within the past $10$~My interval of time. In order to infer a possible range of semimajor axis drift rates $\dot{a}$ for each of the asteroids (except 832 Karin), we assigned them a size $D$ by either (i) directly obtaining the information from the database provided by WISE \citep{mas2011} or (ii) assuming the objects had a $p_V=0.23$ geometric albedo, the median value for the Koronis family as determined from WISE observations \citep[e.g.][]{mas2013}. There were $24$ cases in the first group. 

We then used the linearized theory of heat conduction for spherical bodies from \citet{vok1998} to estimate the maximum possible values of $\dot{a}_{\rm max}$. Note that for an individual body, $\dot{a}_{\rm max}$ also depends on the rotation period $P$ and the thermal inertia of the surface $\Gamma$. Given that we do not know these values for our candidate objects at this time, we assumed $\dot{a}_{\rm max}$ reached its maximum over all possible values of $(P,\Gamma)$. This calculation yielded $\dot{a}_{\rm max}=\pm 3.1\times 10^{-4}/D$ au~My$^{-1}$, where $D$ is in kilometers and the positive and negative values are for the two extreme obliquities of $0^\circ$ or $180^\circ$, respectively. For each object in the sample of $420$ Karin-family candidates, we considered $\dot{a}$ within these limits. 

Proceeding with $0.01$~My timesteps in $T$, we found the minimum possible values of $\Delta\Omega_j(T)$ and $\Delta\varpi_j(T)$ and subsequently calculated $\cC(T)$ for the entire sample. At the epoch corresponding to the $\cC(T)$ minimum, we evaluated the contributions of each asteroid in the sample and identified those with contributions exceeding four times the minimum value as interlopers. These non-converging objects, defined as interlopers, were subsequently removed from the sample. The process was iteratively repeated until no additional interlopers were detected. This refinement resulted in a final selection of $370$ Karin family members that are smaller than (832)~Karin but larger than $1.3$~km (namely $H\leq 16.7$ from our initial selection). The number of removed objects, $49$ out of $420$ (832~Karin not counted), roughly informs us about the fraction of interlopers in the Karin family, namely $\simeq 11$\% (most of which come from the Koronis2 and Kutaissi clusters).

\begin{figure}[t!]
 \begin{center}
  \includegraphics[width=0.47\textwidth]{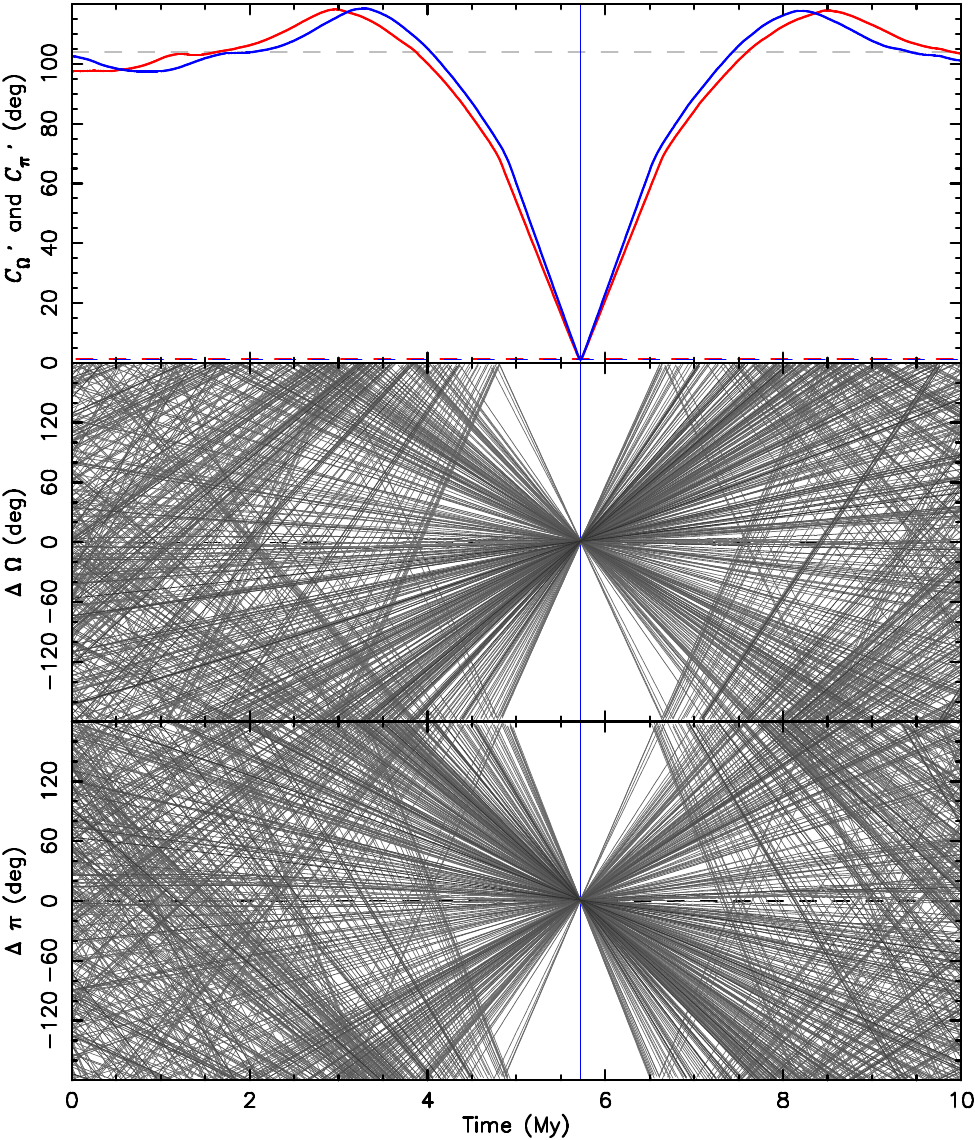}
 \end{center}
  \caption{Convergence of the proper longitude of node (middle panel) and perihelion (bottom panel) of $370$ Karin family members with $H\leq 16.7$ at $T=5.72$~My. Darker lines for larger members. The top panel shows behavior of $\cC^\prime_\Omega$ (red) and $\cC^\prime_\varpi$ (blue) defined in (\ref{conv1}). Both show sharp minima at nearly the same epoch, as required by a satisfactory global solution of $\cC$, attaining the best-fit values $1.29^\circ$ for nodes and $1.05^\circ$ for perihelia. The gray dashed level at the top panel corresponds to $\cC^\prime$ value of a uniform distribution for comparison.}
 \label{fig1}
\end{figure}
\begin{figure}[t!]
 \begin{center}
  \includegraphics[width=0.47\textwidth]{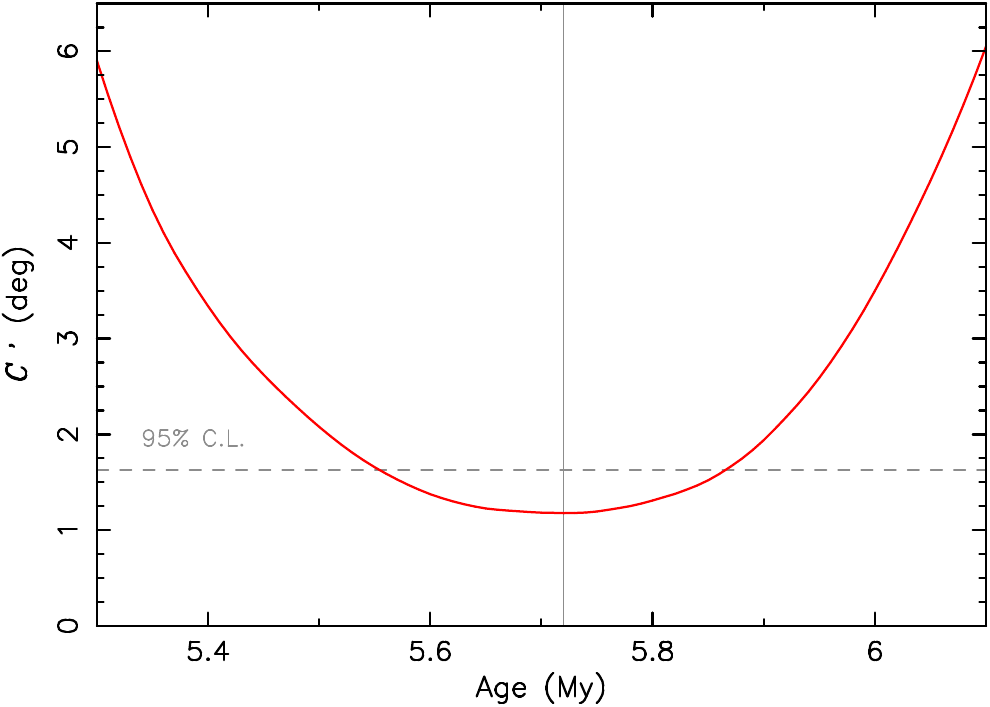}
 \end{center}
  \caption{Behavior of the target function $\cC^\prime$ near its minimum at $T=5.72$~My (vertical gray line) combining convergence of proper nodes and perihelia of $370$ Karin family members with $H\leq 16.7$. The dashed gray horizontal line indicates 95\% confidence level of the age solution assuming $\cC^\prime$ represents a merit function of minimization procedure with $\simeq 1.5^\circ$ uncertainty of the $\Delta \Omega_j$ and $\Delta\varpi_j$ values \cite[each characterized by normal distribution; see Chap.~14 and 15 in][]{nr2007}.}
 \label{fig2}
\end{figure}
\begin{figure}[t!]
 \begin{center}
  \includegraphics[width=0.47\textwidth]{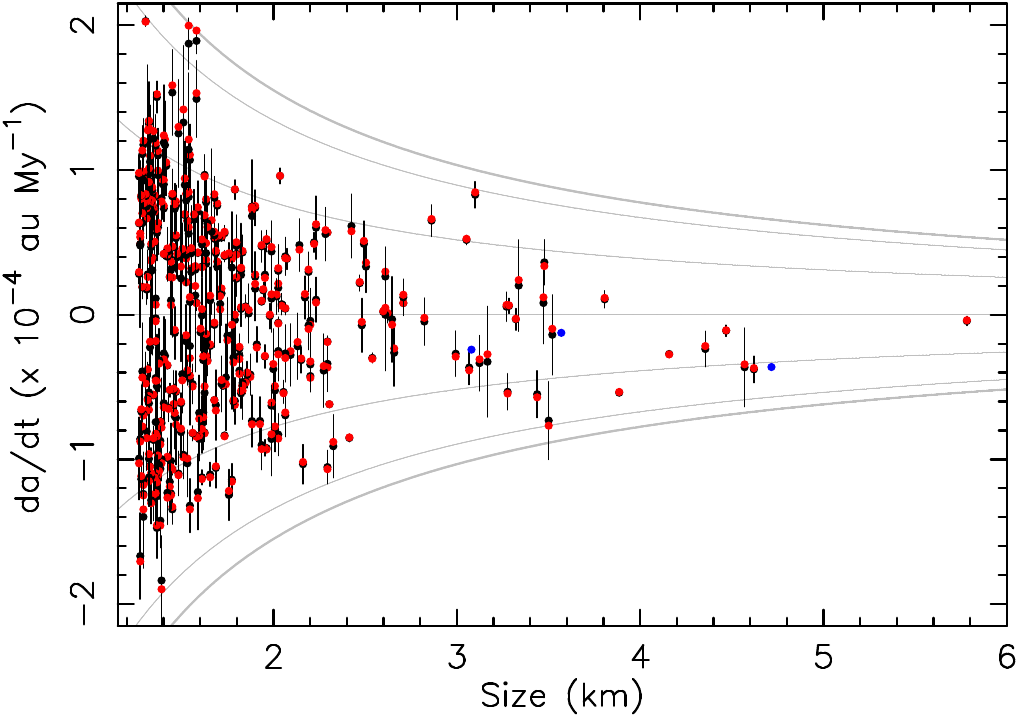}
 \end{center}
  \caption{Adjusted values of mean semimajor axis drift-rate values $\dot{a}$ for $367$ Karin family members with $H\leq 16.7$ as a function of their size $D$. Red symbols correspond to the solution with minimum dispersion of the secular angles expressed by $\cC$ function in (\ref{tf}). The black symbols, and vertical bars, are the mean and standard deviation of their distribution from solutions with $T$ sampling the 95\% C.L. interval of Karin family ages, namely between $5.53$~My and $5.88$~My (Fig.~\ref{fig2}). The enveloping thick gray lines are the estimated maximum values $\dot{a}_{\rm max}=\pm 3.1\times 10^{-4}/D$ au~My$^{-1}$ from the linearized heat conduction model on spherical bodies and bulk density of $2.5$ g~cm$^{-3}$ \citep[e.g.,][]{vok1998}. The positive and negative maximum values correspond to $0^\circ$ and $180^\circ$ obliquity. The thin gray lines are for intermediate obliquity values incremented by $30^\circ$. For sake of completeness, we also displayed the $\dot{a}$ values corresponding to 7719, 33143 and 40510 (blue symbols), whose membership in Karin or Koronis2 families is ambiguous (see Sec.~\ref{karin}).}
 \label{fig3}
\end{figure}
\begin{figure}[t!]
 \begin{center}
  \includegraphics[width=0.47\textwidth]{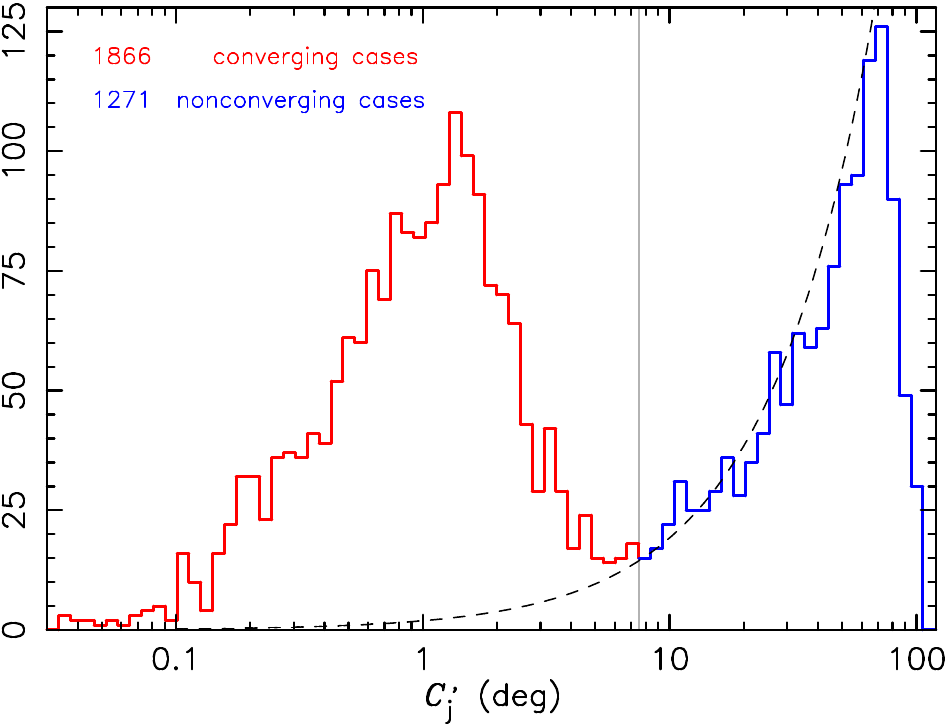}
 \end{center}
  \caption{Differential distribution of $\cC^\prime_j=\sqrt{(\Delta\Omega_j^2+\Delta\varpi_j^2)/2}$ values by which each of the $3217$ small ($H>16.7$) Karin family candidates contributes to the convergence target function (\ref{tf}) at fixed epoch $T=5.72$~My. The distribution is clearly bimodal with $1866$ inputs having $\cC^\prime_j\leq 7.5^\circ$ which we adopted a limit for satisfactory convergence (red segment); these asteroids define the population of small Karin members. The remaining 1217 objects do not exhibit satisfactory convergence to (832)~Karin at epoch of the Karin family formation (blue segment). Given the logarithmic binning at the abscissa, the dashed line characterizes a uniform distribution of $\cC^\prime_j$.}
 \label{fig4}
\end{figure}
\begin{figure*}[t!]
 \begin{center}
  \includegraphics[width=0.9\textwidth]{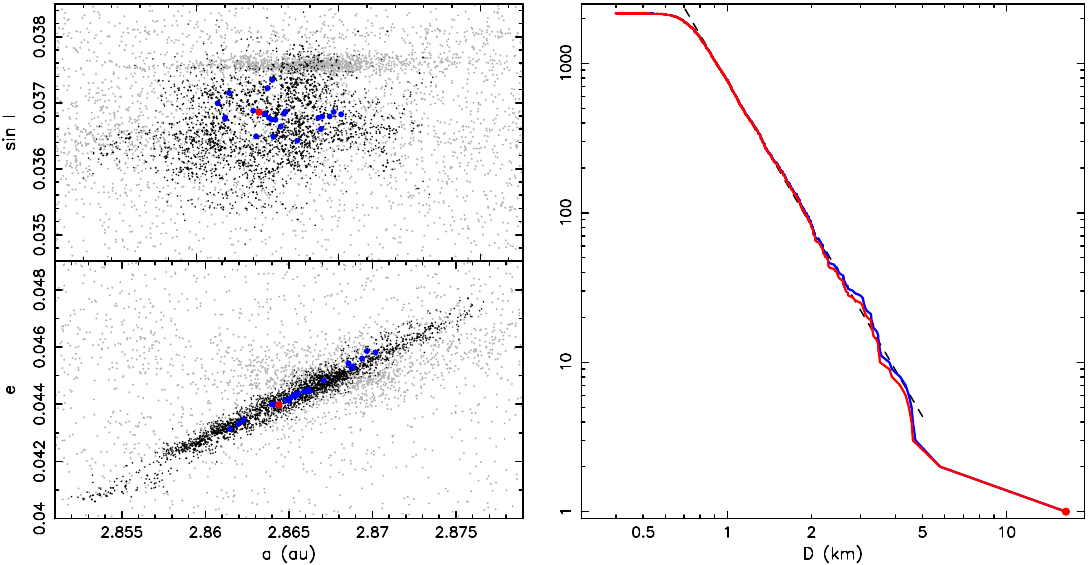}
 \end{center}
  \caption{Left panels: Karin family members (black symbols) projected onto 2D planes of proper semimajor axis $a$ at the abscissa, and proper sine of inclination $\sin I$ (top) and eccentricity $e$ (bottom) at the ordinate. The largest body (832)~Karin shown by red circle, and members with size $\geq 3$~km by blue symbols. The gray dots show unrelated background asteroids in the Koronis family, including the Koronis2 cluster (at $\sin I\simeq 0.0376$). Right panel: Cumulative size distribution for $2161$ members in our conservative identification of Karin family (red curve). Sizes of $21$ largest objects from WISE observations, while those for the remaining members determined from their absolute magnitude $H$ and assumed albedo $p_V=0.23$, a mean value of the Koronis family from WISE data as well \citep{mas2011,mas2013}. In between $0.85$ and $3$~km the distribution is fairly well matched by a power-law $N(>D)\propto D^{-\alpha}$ with a slope index $\alpha=3.2$ (dashed line). For sake of comparison we also show distribution that would include also asteroids 7719, 33143 and 40510 with an ambiguous membership in either Karin or Koronis2 families (see Sec.~\ref{karin}; blue curve).}
 \label{fig5}
\end{figure*}
\begin{figure}[t!]
 \begin{center}
  \includegraphics[width=0.47\textwidth]{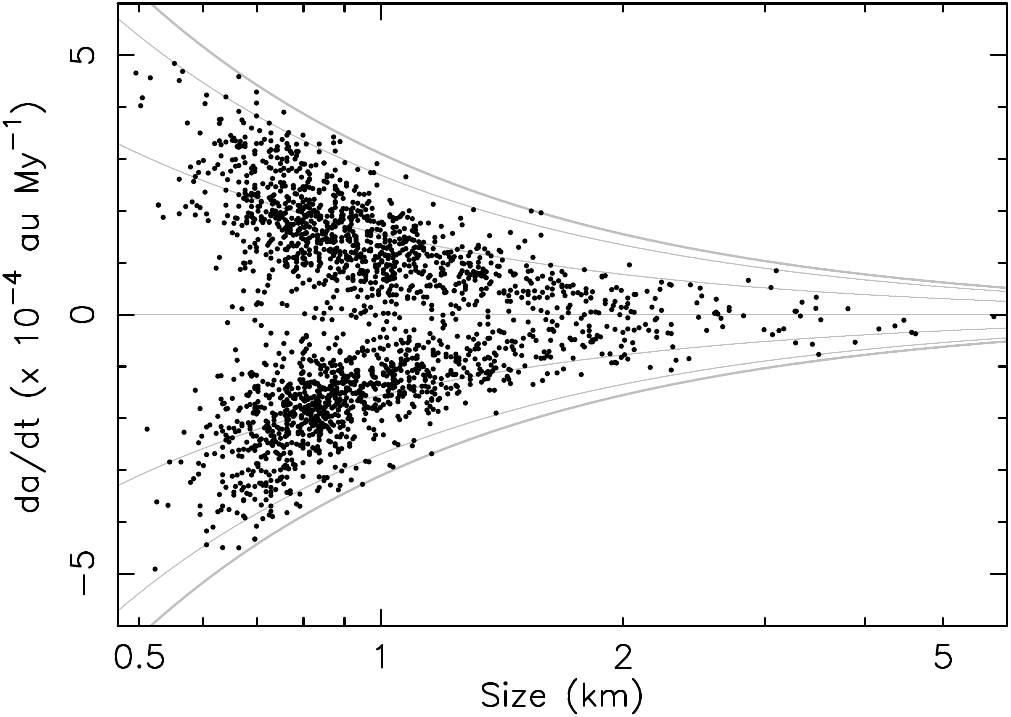}
 \end{center}
  \caption{Mean values of the semimajor axis drift $\dot{a}$ (black symbols) of all $2160$ members in the Karin family (except 832 Karin itself) inferred from convergence of their proper secular angles $\Omega$ and $\varpi$ as a function of their size $D$. The gray lines as in Fig.~\ref{fig3}. Below $\simeq 1.5$~km size, the $\dot{a}$ values have a nonuniform distribution, avoiding zero value and having peaks at about half of the maximum value for both the prograde- and retrograde-rotation regime. This is due to due to the YORP effect that made rotation pole directions of sub-kilometer Karin members tilt away of the ecliptic plane \citep[see][]{car2016}.}
 \label{fig6}
\end{figure}

Figure~\ref{fig1} shows the successful convergence of proper nodes and perihelia toward the orbit of (832)~Karin, while the middle panel shows $\Delta\Omega$ values and the lower panel shows $\Delta\varpi$ values. To verify that the convergence of both elements occurs at the same epoch, we split $\cC^\prime(T)$ into the two components $\cC^\prime_\Omega(T)$ (red curve) and $\cC^\prime_\varpi(T)$ (blue curve) in the top panel. Their minimum values, $1.29^\circ$ for the nodes and $1.05^\circ$ for the perihelia, indeed occur simultaneously at $T=5.72$~My. 

To analyze the behavior of the composite function $\cC^\prime(T)$ near its minimum and to estimate the age of the Karin family using our method, we consider the information shown in Fig.~\ref{fig2}. At first sight, an inconsistency appears between $\cC^\prime_\Omega(T)$ and $\cC^\prime_\varpi(T)$ in the top panel of Fig.~\ref{fig1} and substantially shallower-in-time $\cC^\prime(T)$ in Fig.~\ref{fig2}. In Fig.~\ref{fig1}, however, we considered fixed values of $\dot{a}_j$, namely those that resulted in the best convergence solution, and hold them to compute $\Delta\Omega$ and $\Delta\varpi$ in other epochs. In contrast, in Fig.~\ref{fig2}, at each epoch $T$, we allowed the values $\dot{a}_j$ to vary so that the smallest value $\cC^\prime(T)$ is locally reached. These values are different at each time. This indicates that a much better convergence is achieved over a broader range of epochs near the global minimum at $T=5.72$~My in Fig.~\ref{fig2} than in Fig.~\ref{fig1}. 

To establish a limit on the individual values within $\cC^\prime(T)$ that could provide a confidence interval for the possible age of the Karin family, we formally assume an uncertainty of approximately $\simeq 1.5^\circ$ in both $\Delta\Omega_j$ and $\Delta\varpi_j$. This level reflects the previously mentioned ejection velocities of large Karin fragments, as well as the noise introduced by perturbations from massive bodies in the main belt (see Appendix~\ref{kamean} and Fig.~\ref{figcpv}).

Assuming a Gaussian distribution for both angles and considering $370$ degrees of freedom by adjusting the values of $\dot{a}_j$ for each asteroid, we can set a $\cC^\prime(T)$ threshold for every chosen confidence limit (C.L.) in the age solution \citep[see discussion in Chap.~14 and 15 in][]{nr2007}. For $95$\% C.L., we obtained the value shown by the dashed horizontal line in Fig.~\ref{fig2}. This translates to the Karin age of $T=5.72_{-0.19}^{+0.16}$~My based on our method.

Before proceeding further, we took a closer look at the largest asteroids within the converging sample. We found three bodies, (7719) 1997~GT36, with $D=4.7\pm 0.5$~km, (33143) 1998~DJ7, with $D=3.6\pm 0.7$~km, and (40510) 1999~RU87, with $D\simeq 3.1$~km, all of which have $\sin I > 0.0374$. While the Karin family stretches over a large set of proper inclination values, a distinct feature of the Koronis2 family is its tight confinement in proper inclination values between $0.0374$ and $0.0378$ \citep[e.g.,][]{molnar2009,broz2024}. Interestingly, while these three asteroids converge to (832)~Karin, they also converge to (158)~Koronis as shown in Sec.~\ref{koronis2}. Unlike the other objects in the initial sample of bodies with $H\leq 16.7$ selected in this section, their convergence to either of the two largest remnants in the Karin and Koronis2 families does not provide definitive evidence regarding their membership in one family or the other. In this situation, we adopted a conservative approach and excluded them from the Karin family (though we acknowledge that readers holding an opposing view may prefer to retain them as Karin family members). As a result, we are left with $367$ members in the Karin family with $H\leq 16.7$ (in addition to (832)~Karin itself).

The remarkable one-degree level of convergence in both nodes and perihelia, as demonstrated above, stems from the adjustment of the mean drift rates $\dot{a}_j$ for each body. Figure~\ref{fig3} displays these values for all 367 converging asteroids in our nominal Karin family, along with the three discarded asteroids (7719, 33143, and 40510), plotted as a function of their estimated size $D$: the red symbols represent their solutions corresponding to the minimum $\cC^\prime(T)$ value at $T=5.72$~My, and (ii) the black symbols, with associated uncertainty intervals, represent their mean and standard deviation derived from individual best-fitting convergences with $T$ across the $95$\% C.L. interval of Karin age values (Fig.~\ref{fig2}).

We may conclude that the $\dot{a}_j$ solution is robust, as the variations in their values are relatively small, and the best-fitting values consistently remain close to the mean obtained by sampling the uncertainty in the Karin age. Additionally, all $\dot{a}_j$ values fall within the range defined by $\dot{a}_{\rm max}$ for the estimated size. This indicates that that the allowed interval for $\dot{a}$ was even more generous than necessary. For Karin members with sizes between $\simeq 1.3$ and $3.5$~km, the $\dot{a}$ values are evenly distributed between positive and negative values, pointing to an isotropic distribution of their rotation poles. 

Conversely, it has been previously observed that members with $D\geq 3.5$~km members consistently exhibit negative $\dot{a}$ values, indicating a predominantly retrograde sense of rotation \citep[as noted in][]{nes2004,car2016}. The hypothesis that they all have such values can be tested through photometric observations that can directly determine their rotation states. Such work could also provide important constraints for numerical models describing the disruption of the Karin family’s parent body.

\begin{figure*}[t!]
 \begin{center}
  \includegraphics[width=0.9\textwidth]{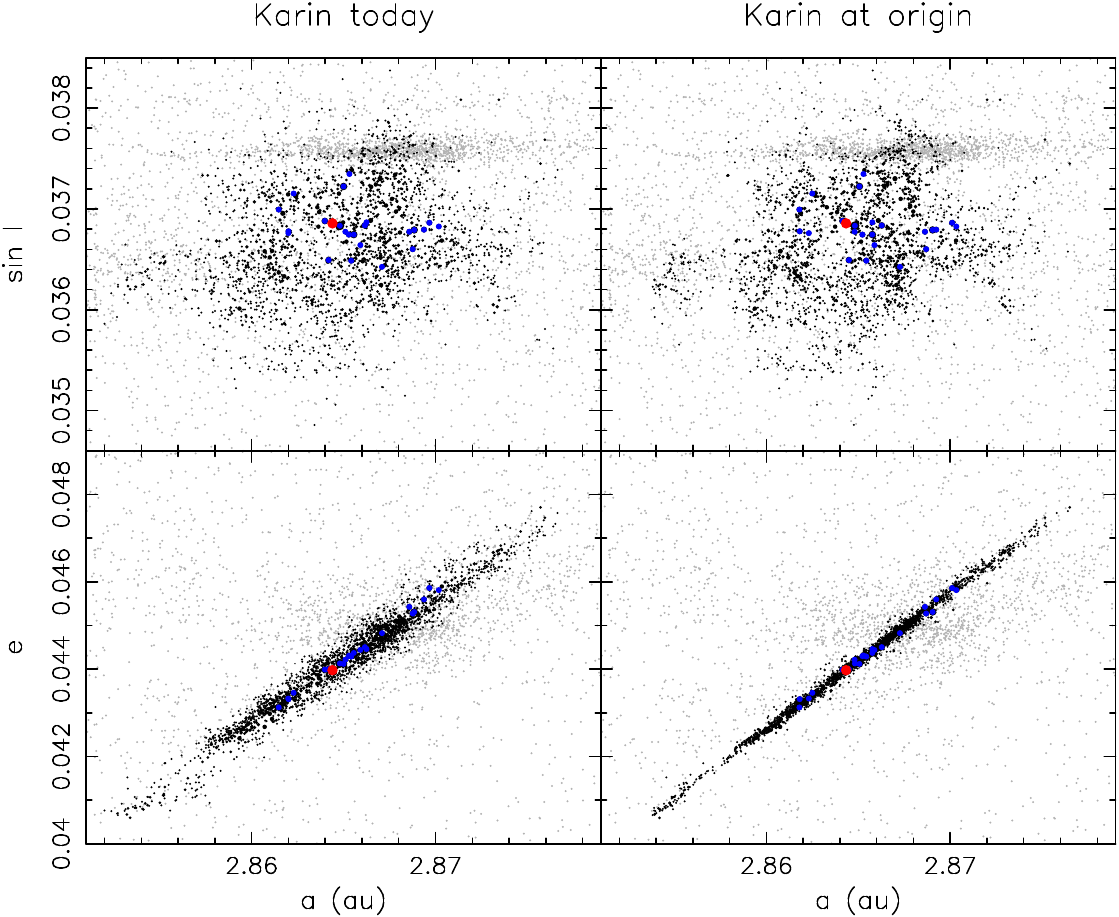}
 \end{center}
  \caption{Karin family projected onto 2D planes of proper semimajor axis $a$ at the abscissa, and proper sine of inclination $\sin I$ (top) and eccentricity $e$ (bottom) at the ordinate. The largest body (832)~Karin shown by red circle, and members with size $\geq 3$~km by blue symbols. The gray dots show unrelated background asteroids in the Koronis family. Left panels show the present-day configuration, right panels is our reconstructed family at its formation $T=5.72$~My ago. For that we kept the $\sin I$ and $e$ values constant, but shifted $a$ values by $-\dot{a}\,T$, with drift rate $\dot{a}$ inferred from convergence of proper secular angles $\Omega$ and $\varpi$. The initial family shows two distinct features: (i) a more confined correlation in the $(a,e)$ plane, and (ii) more focused filamentary structures in the $(a,\sin I)$ plane \citep[both predicted by][using SPH impact simulations]{nes2006}.}
 \label{fig7}
\end{figure*}

\subsubsection{Step~2: Analysis of small member population}\label{ka2}
In the next step, we consider the sample of $3217$ asteroids with $H>16.7$ in our original Karin cluster identified using HCM in proper element space. The set of Karin members with $H\leq 16.7$ precisely determines the family's age. We therefore fixed the age at $T=5.72$~My and focused on which of the $3217$ small asteroids have the potential to converge to (832)~Karin at the selected epoch in the past.  

We use the same convergence method as described earlier, assigning each small asteroid a $\dot{a}_j$ value within the range of $\dot{a}_{\rm max}=\pm 3.1\times 10^{-4}/D$ au~My$^{-1}$. The only difference is that we use a slightly wider range due to the smaller sizes of the asteroids in the second candidate sample. The diameter $D$ was estimated using the absolute magnitude $H$ and a mean geometric albedo of $p_V=0.23$. We did not follow an iterative process, as we anticipated that many of the objects in this sample would not converge with (832)~Karin.

For each asteroid tested, we only optimized the value of $\dot{a}_j$ to reach the best convergence result, which we quantitatively express using a contribution ${\cC^\prime_j}^2=(\Delta\Omega_j^2+\Delta\varpi_j^2)/2$ in $\cC^2$. We set the threshold for acceptable convergence at $\cC^\prime_j\leq 7.5^\circ$.  We argue that this value provides a generous margin, considering the excellent level of convergence observed in the sample of larger Karin members.

Figure~\ref{fig4} shows distribution of the resulting $\cC^\prime_j$ values for the entire sample. As expected, it has a bimodal nature which allows us to separate small members of the Karin family (red segment of the distribution) from interlopers (many of which will contribute to the population of the Koronis2 family; blue segment of the distribution). The division of the two groups appears very close to our choice $\cC^\prime_j\leq 7.5^\circ$, which we consider to be a nice a posteriori justification for our assumption. 

The median value of $\cC^\prime_j$ in the converging group is $1.02^\circ$, which is consistent with the value characterizing the convergence of the larger members of the Karin family (Sec.~\ref{ka1}). The distribution of the $\cC^\prime_j$ values of Karin-unrelated asteroids appears to peak near $70^\circ$, but but this is merely an artifact caused by the use of logarithmic binning in Fig.~\ref{fig4}. In reality, their distribution is uniform, as illustrated by the dashed line. When extrapolated to $\cC^\prime_j\leq 7.5^\circ$, along with the presence of a tail of formally converging objects nearing the arbitrarily chosen separation threshold, this suggests that some asteroids in this group may still be interlopers from the background population or the Koronis2 family. We estimate that there could be several dozen to about a hundred such interlopers.

To investigate further, we closely examined the orbits contributing to the 'neck' region between the converging and non-converging groups. We identified 73 asteroids that, based on their proper orbital elements, reside within a halo that is too distant from the core members of the Karin family. One particularly suspicious feature is their alignment in $\sin I$ near $0.0375$, a value corresponding to the center of the Koronis2 family. For this reason, we decided to exclude these 73 asteroids from the Karin family. This refinement leaves us with $1,793$ confirmed Karin members with $H>16.7$. Combined with the sample identified earlier in Sec.~\ref{ka1}, the updated population of Karin family members, including (832) Karin, now totals $2,161$ asteroids.

The left panels in Fig.~\ref{fig5} show the distribution of the identified members of the Karin family in the two dimentional  projections of proper orbital elements: (i) $(a,\sin I)$ (top), (ii) $(a,e)$ (bottom). A highly correlated diagonal distribution in the $(a,e)$ projection, which is a distinct characteristic that played a key role in the family's initial discovery \citep{nes2002}, strongly supports the idea that the parent body of the family was disrupted near the perihelion of its orbit.  From a quantitative perspective, $|f_\star|\leq 30^\circ$ from Eqs.~(\ref{gau1}) and (\ref{gau2}), where $f_\star$ is the true anomaly of the parent body at the collision.

The right panel of Fig.~\ref{fig5} shows the cumulative size frequency distribution (SFD) $N(>D)$ of the Karin family. 
Only $21$ of the largest members have their sizes determined based on the analysis of WISE data. For the remaining asteroids, we assumed a geometric albedo of $p_V=0.23$, the average value for the surrounding Koronis family \citep[e.g.,][]{mas2013}. The population of the largest members is too small to enable a meaningful fit of $N(>D)$ with a sufficiently smooth and simple function. However, within the size range of approximately $0.8$ and $2.5$~km, a power law approximation $D(>D)\simeq D^{-\alpha}$ with $\alpha=3.20\pm 0.01$ provides an excellent fit. Below $0.8$~km, the population is observational incomplete. 

We verified the $0.8$~km threshold using well-characterized observations of the Catalina Sky Survey between 2016 and 2022, analyzed extensively by \citet{neomod2}. This led us to infer the completeness level as $H\simeq 17.2$. With our assumed albedo value, this translates to a size of approximately $1$~km. This data set presents an opportunity to revisit the modeling efforts of \citet{nes2006} to gain a more detailed understanding of the disruption of the Karin family's parent body. For example, we may be able to more accurately constrain the size of the parent body, or we may now be able to better estimate the unseen mass of fragments below the completeness limit. Either advance could yield highly valuable insights (see the discussion in Sec.~\ref{concl}).

Figure~\ref{fig6} shows the adjusted drift rates $\dot{a}_j$ for the entire population of $2,161$ Karin family members. The data on small members presented in this section introduces notable differences compared to Fig.~\ref{fig3}. The information does not just change the number of data points. It instead also reveals the onset of new behavior for sizes $D\leq 1.1$~km. While for larger sizes the $\dot{a}_j$ values are approximately evenly distributed across the entire admissible range of $\pm\dot{a}_{\rm max}$, the distribution for smaller sizes exhibits distinct characteristics: (i) it avoids very small drift rates, and (ii) it shows a preference for drift rates corresponding to rotation pole latitudes of $\pm 30^\circ$. 

This polarization in the drift rates of small Karin members was first identified and explained by \citet{car2016} as a result of the obliquity changes driven by the Yarkovsky-O'Keefe-Radzievskii-Paddack (YORP) effect \citep[for more details on the YORP effect, see, e.g.,][]{vetal2015}. In this work, the trend is confirmed using a much larger sample of asteroids and is further extended to even smaller sizes.

Finally, having achieved a very high level of convergence of the secular proper angles of all Karin members to (832)~Karin at the epoch of the family’s formation, we can now explore what additional details about the initial family configuration might still be uncovered. A straightforward approach is to project the other proper orbital elements backwards in time to their original values. However, there are some inherent limitations to this method. Specifically, we assume the proper eccentricities and inclinations remain constant, meaning their initial values are equal to their current ones. Yet, the proper semimajor axes involve a non-trivial, albeit simplified, adjustment, namely that their initial values are computed as $a_j(T)=a_j(0)-\dot{a}_j\,T$. This effect is small and has not been directly used so far. 

The changes in the Karin family configuration within the 3D space of proper orbital elements $(a,e,\sin I)$ are displayed in Fig.~\ref{fig7}. The present-day distribution is shown in the left panels, while the reconstruction of the initial configuration appears in the right panels. Interestingly, the differences are both significant and entirely consistent with expectations. In the $(a,e)$ plane, the initial distribution becomes even more tightly aligned with the linear correlation. This further constrains the possible range of the parent body's true anomaly, $f_\star$ at the epoch of family formation, being very close to the value $f_\star=0^\circ$. Similarly, notable changes are evident in the $(a,\sin I)$ projection, where the Karin family fragments begin to form a filamentary structure, resembling a sharper, more focused version of the currently diffuse distribution.

Even more intriguingly, numerical simulations of the Karin family's breakup performed by \citet{nes2006} predict the existence of such filamentary structures. This agreement between the theoretical predictions and our data-driven reconstruction appears to validate both approaches. Bridging the remaining gaps between the two methodologies represents an exciting direction for future work. Achieving this will likely require incorporating the past orbital dynamics of Karin family members, extending beyond the simplified model based solely on proper elements.

\subsubsection{Step~3: Further improvements?}\label{ka3}
The convergence of secular angles has been used as a standard method for estimating the age of very young asteroid families since the work of \citet{nes2002}. This approach, however, only represents a necessary condition for reconstructing the family configuration at its origin. The ultimate goal would be to align the full heliocentric state vectors of all family members together, namely at their mutual distance defined by the Hill radius of the parent body and very small relative velocities. Yet, for families consisting of thousands of members and ages exceeding millions of years, achieving this level of reconstruction appears to be an unattainable task.  So far, it has only been successfully achieved in the case of much younger asteroid pairs \citep[e.g.,][]{vok2008}.

Despite these challenges, it may still be possible to improve upon this method when applied to asteroid families such as the Karin family. In this section, we explore this possibility. Fully leveraging this novel approach necessitates transitioning from proper to mean orbital elements and replacing analytical techniques with numerical methods. The initial steps of this new methodology are discussed in greater detail in Appendix~\ref{kamean}.

The initial ejection velocities of Karin family members, as referenced to the largest remnant (832)~Karin, may be expressed in its orbital plane using the radial, transverse and out-of-plane components $(V_r,V_t,V_z)$. To map these values into differences in orbital elements, specifically, the osculating orbital elements, we must also determine the argument of perihelion $\omega_\star$ and the true anomaly $f_\star$ of (832)~Karin at the time of origin. Moreover, the resulting mathematical expressions become more straightforward by scaling the components $(V_r,V_t,V_z)$ by the heliocentric orbital velocity $V_0\simeq 17,600$ m~s$^{-1}$, which corresponds to a circular orbit with a radius equal to the proper semimajor axis of (832)~Karin. These non-dimensional velocity components are denoted $(v_r,v_t,v_z)$. 

Neglecting first- and higher-order eccentricity and inclination terms, the Gauss equations \citep[e.g.,][]{bfv2003} describe for each fragment the map between $(v_r,v_t,v_z;\omega_\star,f_\star)$ and $(\delta a,\delta e,\delta \sin I,\delta \Omega,\delta\varpi)$.
This correspondence may be conveniently broken into three groups:
\begin{itemize}
\item the transverse component $v_t$ of the ejection velocity is directly connected with the change in semimajor axis $\delta a$ as
\begin{equation}
 \frac{\delta a}{a} = 2\,v_t\; , \label{gau1}
\end{equation}
\item a linear combination of the transverse $v_t$ and radial $v_r$ components of the ejection velocity determines changes in eccentricity $\delta e$ and in longitude of perihelion $\delta\varpi$ expressed as a simple 2D-rotation by $f_\star$
\begin{equation}
 \left(\begin{array}{c} \delta e \\ e \,\delta\varpi \end{array}\right) = \left( \begin{array}{cc} \cos f_\star &  -\sin f_\star \\ \sin f_\star & \cos f_\star \end{array} \right)\;
  \left(\begin{array}{c} 2\,v_t \\ -v_r \end{array}\right)\; , \label{gau2}
\end{equation}
\item the out-of-plane velocity component $v_z$ corresponds to the change in inclination $\delta \sin I$ and longitude of node $\delta \Omega$. It also has a straightforward interpretation in a 2D plane. Specifically, the pair $(\delta\sin I,\sin I \delta\Omega)$ plays the role of the Cartesian axes, while the out-of-plane velocity $v_z$ and angle $\omega_\star+f_\star$ function as polar coordinates
\begin{eqnarray}
 \delta \sin I &=& v_z\,\cos\left(\omega_\star+f_\star\right)\; , \label{gau3a} \\
 \sin I\,\delta \Omega &=& v_z\,\sin\left(\omega_\star+f_\star\right)\; . \label{gau3b} 
\end{eqnarray}
\end{itemize}
In fact, an even simpler form of the velocity-to-orbital element relationships can be obtained by using non-singular orbital elements
$k=e\cos\varpi$, $h=e\sin\varpi$, $q=\sin I\cos\Omega$ and $p=\sin I\sin\Omega$, or their combination in complex variables ${\bf z}=k+\imath\, h$ and $\boldsymbol{\zeta}=q+\imath\,p$. Note that $({\bf z},\boldsymbol{\zeta})$, in replacement of $(e,I,\Omega,\varpi)$, are (i) more appropriate for small eccentricity and inclination orbits and (ii) directly related to the construction of the proper orbital elements \citep[e.g.,][]{km2000,ketal2003,nes2024}. 

Interestingly, the map between the variation of the orbital elements $(\delta a,\delta {\bf z},\delta\boldsymbol{\zeta})$ and the imparted relative velocity $(v_r,v_t,v_z)$ requires only a single parameter $F_\star=\varpi_\star + f_\star$, namely the true longitude in orbit of the parent body at the epoch of family formation. While Eq.~(\ref{gau1}) still holds, the remaining system of equations is simplified to
\begin{eqnarray}
 \delta {\bf z} &=& \left(2v_t-\imath\, v_r\right)\, \exp\left(\imath F_\star\right)\; , \label{gau4}\\
 \delta \boldsymbol{\zeta}&=& v_z\, \exp\left(\imath F_\star\right)\; , \label{gau5}
\end{eqnarray}
still valid to the zero order in orbital eccentricity and inclination.

We now aim to proceed in the opposite direction, namely, from differences in orbital elements to determining ejection velocities. Knowing the values of $(\delta a,\delta e,\delta \sin I,\delta \Omega,\delta\varpi)$ or equivalently $(\delta a,\delta {\bf z},\delta\boldsymbol{\zeta})$, we could --in an ideal world-- use Eq.~(\ref{gau1}), (\ref{gau4}) and (\ref{gau5}) to infer the relative velocity with which the body separated from (832)~Karin. It is also worth noting that the angle $F_\star$ is identical for all fragments, which should further simplify the procedure.

For example, the real and imaginary parts of $\delta\boldsymbol{\zeta}$, which we can construct from $\delta \sin I$ and $\sin I\delta\Omega$ of all family members, should fall along a single line in the respective complex plane with a phase given by $F_\star$. Similarly, with $v_t$ components determined from the values of $\delta a$, the consistency of the configuration requires that the values of $\delta{\bf z}$ must map onto $2v_t-\imath\,v_r$ using a simple rotation by the angle $-F_\star$ (Eq.~\ref{gau5}). Along with identifying the values of $2v_t$ with $\delta a/a$ (Eq.~\ref{gau1}), this transformation reveals the radial components $v_r$.
\begin{figure*}[t!]
 \begin{center}
  \includegraphics[width=0.9\textwidth]{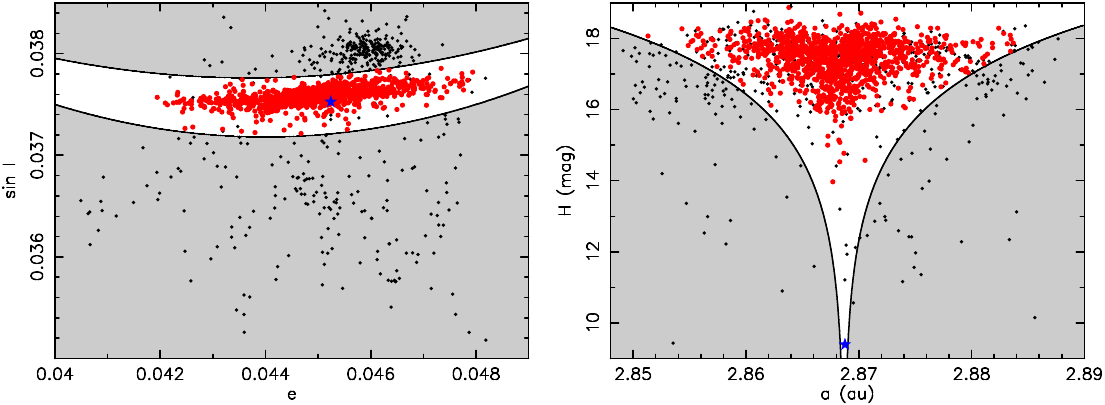}
 \end{center}
  \caption{Pre-selection of Koronis2 family candidate members. Black sysmbols are all $1702$ asteroids that remained in the HCM-identified Karin/Koronis2 family after Karin members were removed. A distinct property of Koronis2 real candidates, shown by red symbols, is a tight confinement of their proper inclination values about $\sin I\simeq 0.0376$ (left panel). We used the two quadratic curves in the $(e,\sin I)$ 2D space to characterize the zone of the Koronis2 family (bodies in the regions were eliminated). Asteroids at lower values of $\sin I$ are too far away in proper element space (median HCM distance from (158)~Koronis of $60$ m~s$^{-1}$) and we consider them background members of the Koronis family. The tight cluster of asteroids at proper $e\simeq 0.0455$ and $\sin I\simeq 0.038$ are members of the interloping Koronis3 family \citep[see Fig.~1 in][]{broz2024} \citep[alternatively named Kutaissi family by][]{nes2025}. The right panel presents the same data in the 2D projection of the $(a,H)$ space. The proper semimajor axes of many large asteroids in the gray zone differ significantly from that of (158)~Koronis (indicated by the blue star).}
 \label{fig8}
\end{figure*}

Although the procedure may seem straightforward, the main challenge lies in the fact that we do not know $(\delta a,\delta {\bf z},\delta\boldsymbol{\zeta})$ of the osculating orbital elements when the family formed. Accordingly, we need to approximate them in some manner. We could use differences in the proper orbital elements $(\Delta a,\Delta {\bf z},\Delta\boldsymbol{\zeta})$ at the moment of their convergence to represent $(\delta a,\delta {\bf z},\delta\boldsymbol{\zeta})$. The issue is that as expected, they do not satisfy conditions (\ref{gau4}) and (\ref{gau5}). Our attempts to adjust some of the parameters, such as the epoch $T$, the Yarkovsky drift $\dot{a}_j$, the proper frequencies%
\footnote{These parameters have formal uncertainties inherent to the procedure used to compute them. Their actual uncertainties, however, may be an order of magnitude larger. For example, test simulations reveal that gravitational perturbations of Karin family member orbits by Ceres, Pallas, and Vesta introduce noise in the proper elements equivalent to initial velocity components of approximately $\simeq 1.5$ m~s$^{-1}$, even over a timescale of just a few million years (see Appendix~\ref{kamean} and Fig.\ref{figcpv}).} 
$g_j$ and $s_j$, and their derivatives with respect the semimajor axis, unfortunately did not lead to a satisfactory solution. 

We believe that the transformation between proper and osculating orbital elements introduces sufficient complexities to make the former unsuitable as substitutes for the latter in the above relations, particularly at the required meter-per-second precision for velocity components. In Appendix~\ref{kamean}, we investigate whether mean orbital elements, determined numerically for a sample of the largest Karin members identified above, might offer greater success in constraining the parameters of the initial velocity field.

\subsection{Koronis2 family}\label{koronis2}
We now return to the population of the originally identified Karin/Koronis2 family, as determined using HCM in the 3D space of proper orbital elements. This cluster consists of 3,863 asteroids, of which 2,161 have been assigned to the Karin family (Sec.\ref{karin}). The remaining 1,702 asteroids represent potential members of the Koronis2 family. Since this family has also been proposed to be young \citep[e.g.,][]{molnar2009, broz2024}, we apply a similar convergence method to identify its members, as previously done for the Karin family, but now considering (158)~Koronis as the largest body.

As with the second step of the Karin family analysis (Sec.~\ref{ka2}), we first exclude obvious or suspected interlopers from the sample. This is necessary because convergence of secular nodes and perihelia alone is too weak of a criterion (any main belt asteroid can eventually converge to (158)~Koronis in these two parameters alone). Membership in the Koronis2 family must also meet additional criteria related to proper element location and size, which essentially ensure that the relative velocity at convergence remains sufficiently small.

The tight clustering of the Koronis2 members in proper inclination proves beneficial in this regard. As a result, we analyzed the distribution of the 1,702 candidate asteroids in the 2D projection of proper orbital elements $e$ and $\sin I$, discarding those located too far from the main Koronis2 cluster (gray regions in the left panel of Fig.~\ref{fig8}). We chose this projection because it provides the best way to identifying members of another interloping family, referred to as Koronis3 or Kutaissi by \citet{broz2024} and \citet{nes2025}: see the distinct cluster at about proper $e\simeq 0.0455$ and $\sin I\simeq 0.0376$. Asteroids on orbits with low values $\sin I$, defied as the bottom gray region in the left panel, are dispersed and their median HCM distance from (158)~Koronis is $\sim 60$ m~s$^{-1}$. We consider them to be interloping objects, possibly members of the Koronis family. 

We also removed several tens of asteroids that were too far from (158)~Koronis in proper $a$ for their absolute magnitude value $H$. This is the gray region shown on the right panel of Fig.~\ref{fig8}, delimited by curves showing $H= H_\star(a)=5\log\left(|a-a_{\rm c}|/c\right)$ with $a_{\rm c}=2.8687$~au and $c=4.5\times 10^{-6}$~au. The ejection speed needed to reach these curves from (158)~Koronis would be $\geq 45$ m~$^{-1}$ \citep[see Sec.~2.3 of][]{vok2006}. 

After completing this second step, the sample of potential Koronis2 members that satisfied the criteria in both the $(e,\sin I)$ and $(a,H)$ spaces was reduced to 1,250 asteroids. These are represented by red symbols in Fig.\ref{fig8}. Admittedly, a small fraction of Koronis2 members near the borders of the elimination lines in Fig.~\ref{fig8} may have been mistakenly excluded, but we believe this fraction is relatively small. 

\begin{figure*}[t!]
 \begin{center}
 \begin{tabular}{ccc}
  \includegraphics[width=0.31\textwidth]{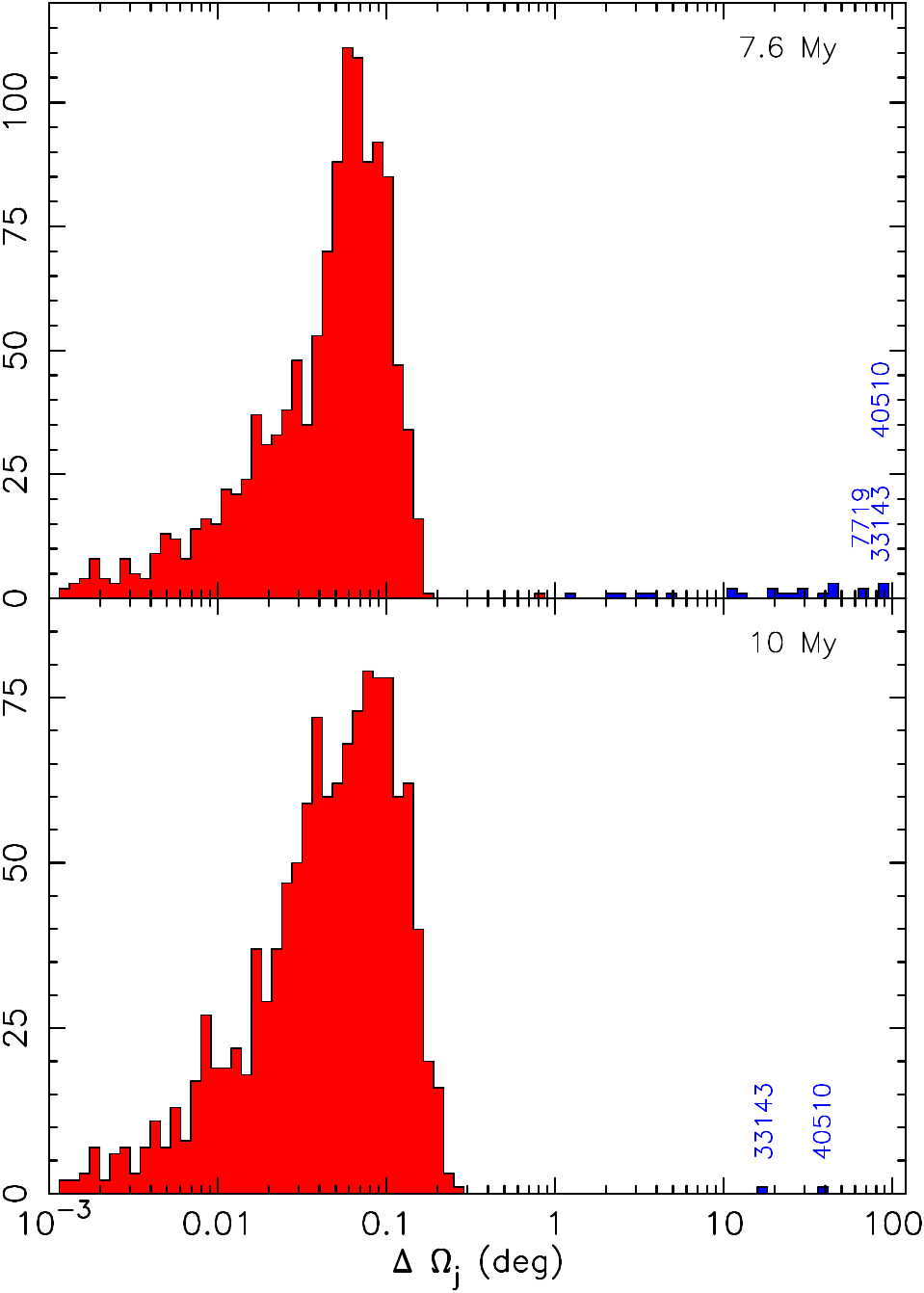} &
  \includegraphics[width=0.31\textwidth]{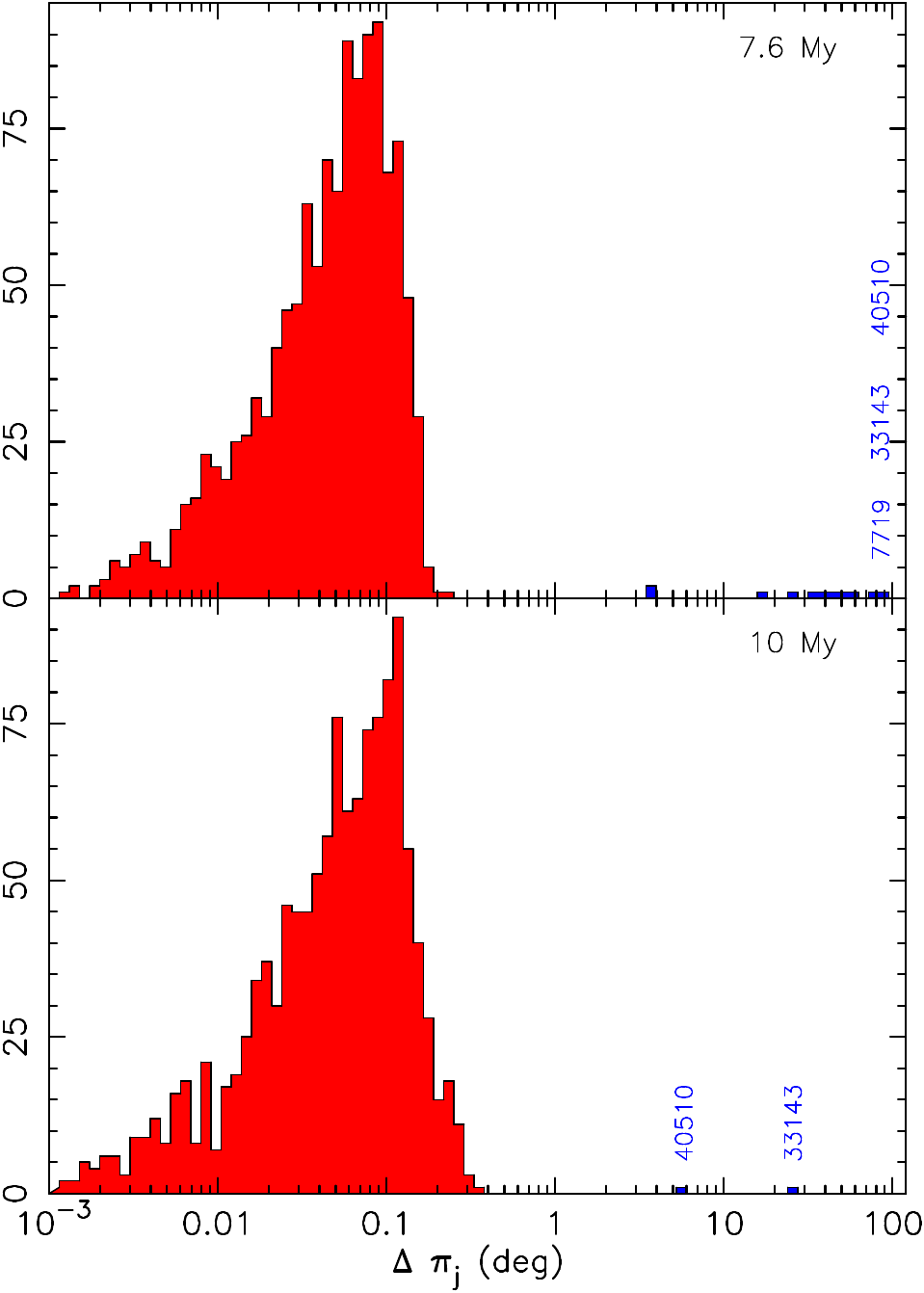} &
  \includegraphics[width=0.31\textwidth]{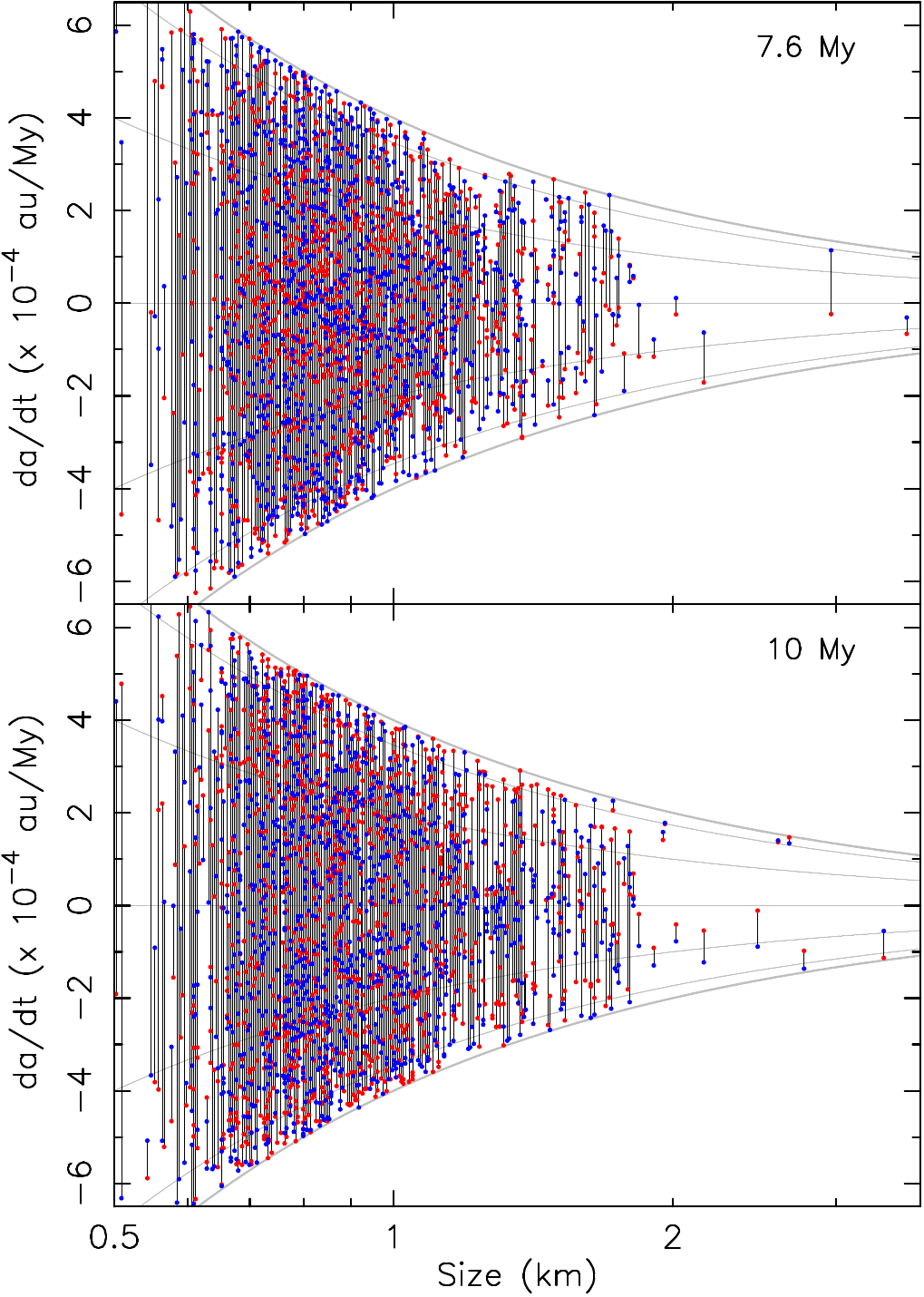} \\
 \end{tabular}
 \end{center}
	\caption{Left and middle panels: Differential distribution of optimum $\Delta\Omega_j$ (left) and $\Delta\varpi_j$ (middle) for $1250$ Koronis2 candidate members at two epochs in the past, $7.6$~My (top) and $10$~My (bottom). Optimization includes only adjustment of the semimajor axis drift rate $(da/dt)_j$ withing the theoretical limit  $\dot{a}_{\rm max}=\pm 4\times 10^{-4}/D$ au~My$^{-1}$. The bulk of mostly small asteroids converges well within the $1^\circ$ threshold (red histogram). A certain number of larger asteroids do not converge since their minimum $\Delta\Omega_j$ and $\Delta\varpi_j$ values remain large (tens of degrees; blue histogram). At $7.6$~My, there are $23$, resp. $11$, such bodies including (7719), (33143) and (40510) that are also Karin family candidates (Sec.~\ref{ka1}). The $D>2$~km asteroids (15501), (55434), and (113021), however, also belong to this non-convergent class. Shifting the epoch of origin to $10$~My, only (33143) and (40510) remain in this class. Right panels: The adjusted $(da/dt)_j$ values for converging Koronis2 candidates at the respective epochs: red symbols for convergence of nodal longitude, blue symbols for convergence of perihelion longitude (the values corresponding to the same asteroid are connected with black interval). The gray lines delimit the accepted drift rate values for a given size $D$ (abscissa, logarithmic scale used as in Fig.~\ref{fig6}) with the characteristic $\propto 1/D$ dependence.}
 \label{fig9}
\end{figure*}

\subsubsection{Nominal convergence of Koronis2 family}
The largest member of the family, (158)~Koronis, is the only asteroid with a sufficient observational dataset to allow determination of its physical parameters. Lightcurve observations and stellar occultations have provided a well-constrained spin model, revealing a rotation period of $14.206$~hr and pole orientation $(30^\circ,-64^\circ)$ in ecliptic longitude and latitude \citep[e.g.,][respectively]{sli2003,dur2011}. 

Additionally, occultation data analyzed in \citep{dur2011} suggest a volume-equivalent size of $38\pm 5$~km. This value is in conflict with the size $47.7\pm 0.6$~km reported by the WISE team, perhaps due to albedo $p_V=0.14\pm 0.01$, unusually low compared to the mean albedo value of the Koronis family \citep[e.g.,][]{mas2011}. Further analyses of WISE’s cryogenic and post-cryogenic phases resulted in somewhat smaller size estimates between approximately $\simeq 31-39$~km \citep{mas2012,mas2014}, while other studies have reported sizes for Koronis within this interval as well \citep[e.g.,][]{ted2002, rw2010, usui2011, ali2018}. Although there is some tension among these size estimates, the implications for our analysis are not particularly significant. 

Here we adopt the $38$~km diameter for (158) Koronis, partially because it is based on occultation data, but also because we suspect the anomalous WISE albedo may have been caused by the object's irregular shape.  This size yields an estimated Yarkovsky drift rate of $\dot{a}\simeq -0.5\times 10^{-5}$ au~My$^{-1}$. This value is low compared to the drift rates of smaller members in the Koronis2 family, for which we conservatively assume $\dot{a}_{\rm max}=4\times 10^{-4}/D$ au~My$^{-1}$. Note this is a slightly larger value than used in the case of the Karin family (Secs.~\ref{ka1} and \ref{ka2}), since we aimed at giving somewhat larger parameter space to the Yarkovsky clones of the Koronis2 members to reach possibility of their convergence.

Following the approach used for the Karin family, we treat the drift rate $\dot{a}$ of (158)~Koronis as fixed, while allowing adjustments for the drift rates of all other candidate members of the family. Although the WISE catalog provides size estimates for five of these smaller members, no additional information about their sizes or spin states is currently available. For the proper frequencies $g_j$ and $s_j$ we use their nominal values and note that the majority of the population has uncertainties smaller than $2\times 10^{-4}$ arcsec~yr$^{-1}$, with only about $5$\% of the population showing larger uncertainties. Analyzing the nominal $g_j$ and $s_j$ values, we also determined the mean values of their gradient with respect to proper $a$, namely $(\partial s/\partial a)=-69.8\pm 0.6$ arcsec~yr$^{-1}$ and $(\partial g/\partial a)=86.9\pm 0.9$ arcsec~yr$^{-1}$.
\begin{figure*}[t!]
 \begin{center}
  \includegraphics[width=0.9\textwidth]{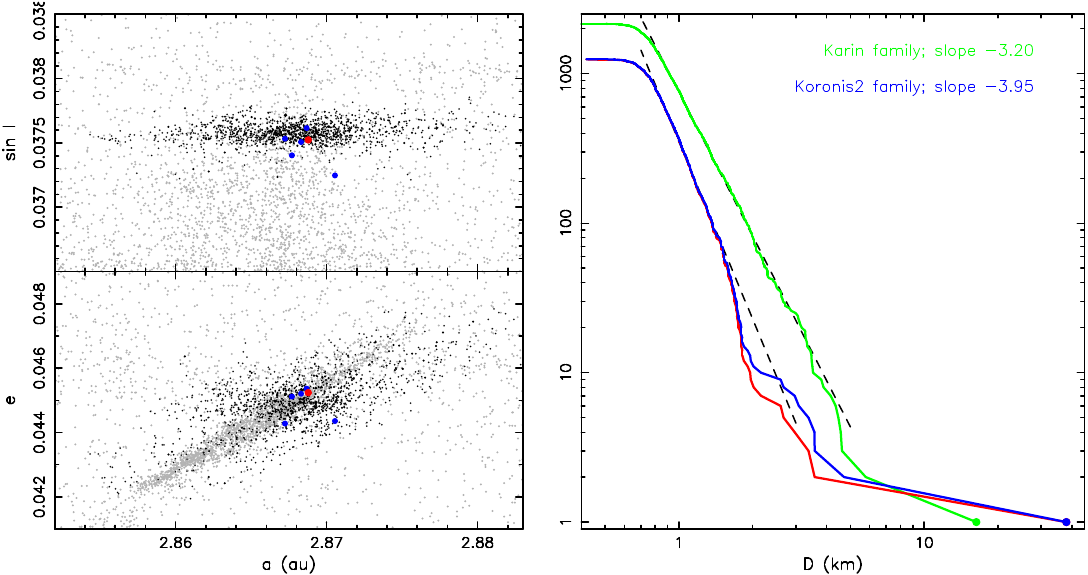}
 \end{center}
  \caption{Left panels: Koronis2 family members (black symbols) projected onto 2D planes of proper semimajor axis $a$ at the abscissa, and proper sine of inclination $\sin I$ (top) and eccentricity $e$ (bottom) at the ordinate. The largest body (158)~Koronis shown by red circle, and members with size $\geq 3$~km by blue symbols. The grey dots show unrelated background asteroids in the (old-)Koronis family, including the Karin cluster. Right panel: Cumulative size distribution for $1250$ members in our full identification of the Koronis2 family (blue) compared to the Karin family (green); the red curve shows a reduced version of Koronis2 without asteroids (7719), (33143) and (40510). The sizes of the $4$ largest objects are from WISE observations, while those for the remaining members were determined from their absolute magnitude $H$ and assumed albedo $p_V=0.23$, the mean albedo value for the Koronis family from WISE data \citep{mas2011,mas2013}. In between $0.8$ and $1.5$~km, the Koronis2 cumulative SFD is reasonably well matched by a power-law $N(>D)\propto D^{-\alpha}$ with a slope index $\alpha=3.95$ (dashed line). This slope is steeper than that of the Karin family over the same size range.}
 \label{fig10}
\end{figure*}

The analysis of the past orbital convergence of Koronis2 members requires a fundamentally different approach than the one previously applied to the Karin family. In the case of the Karin family, even the nominal orbits, without any adjustment of free parameters (i.e., 
$\dot{a}_j=0$ for all fragments), show a reasonable convergence of secular angles at a level of $\cC^\prime\simeq 15^\circ-20^\circ$ level near $T\simeq 5.8$~My \citep[see already][]{nes2002}.
Adjusting $\dot{a}$ serves only to refine the convergence to a higher precision ($\cC^\prime\simeq 1^\circ$), as first demonstrated by \citet{nes2004}, but this process neither significantly alters the epoch $T$ of convergence nor affects its uniqueness.

For the Koronis2 family, however, the situation is markedly different. The nominal orbits of its fragments do not show a clear convergence of nodes and perihelia. Achieving such convergence relies entirely on incorporating nonzero semimajor axis drifts $\dot{a}_j$. Moreover, the maximal drift rates for each Koronis2 candidate member impose a certain minimum value of $T$, the time of convergence in the past, for that particular body. Smaller drift rates, on the other hand, permit larger $T$ values, typically spanning a wide range of past epochs. The intersection of these constraints across all candidate members still results in a substantial uncertainty regarding the family's age.

Complicating matters further, the convergence of the largest asteroids in the candidate list often pushes the epoch $T$ to later times, potentially conflicting with the earlier convergence of smaller candidate asteroids. This discrepancy introduces ambiguity and necessitates subjective decisions about true membership in the Koronis2 family.

To highlight some of the challenges faced in determining the age of the Koronis2 family, we begin by testing the convergence of proper node and perihelion values independently. In our first trial, we looked at $T$ between $7$ and $15$~My.  We highlight two values: (i) $T=7.6$~My, which was motivated by the work of \citep {broz2024}, and (ii) $T=10$~My (to illustrate the change of the results for older epochs). Our results are shown in Fig.~\ref{fig9}. 

At both epochs shown, the majority of Koronis2 candidate members exhibit convergence of nodes and perihelia to (158)~Koronis for reasonable semimajor axis drift-rate values. Interestingly, however, exceptions primarily occur for larger candidates, including asteroids (7719), (33143), and (40510). These three objects are likely members of the Karin family (Sec.~\ref{ka1}) and thus may not pose a problem for determining the age of the Koronis2 family. Nevertheless, there are other potential Koronis2 members with $D>2$~km, such as (15501), (55434), and (113021), whose proper nodes and perihelia also fail to converge to (158)~Koronis at $T=7.6$~My.  Despite this, their proper 
$(a,e,\sin I)$ orbital elements and absolute magnitudes $H$ place them well within the expected region of the Koronis2 family.

When $T$ is shifted to 10 My, only (33143) and (40510) fail to achieve convergence of nodes and perihelia. Additionally, the panels in the right column of Fig.~\ref{fig9} display the semimajor axis drift rates $\dot{a}_j$ required for the convergence of nodes and perihelia. While these values are often similar, they are not identical. In fact, achieving simultaneous convergence of both nodes and perihelia using the same $\dot{a}_j$ value with our simplified method would require pushing $T$ beyond $14$~My.  
  
Admittedly, determining a potential family age near or beyond 10 My may require more precise methods than simply relying on the convergence of proper secular angles. Small variations in the proper frequencies $g_j$ and/or $s_j$, and particularly in their derivatives with respect to the semimajor axis, $(\partial g/\partial a)_j$ and $(\partial s/\partial a)_j$, which we assumed remain constant throughout the entire Koronis2 orbital region, could suggest that the dynamics are more complex than previously outlined. In order to test the effect of fine-tuning of $(\partial g/\partial a)_j$ and $(\partial s/\partial a)_j$ values, we let them to span a $\pm 1$ arcsec~yr$^{-1}$ interval about the above-mentioned nominal values in the region of the Koronis2 family individually for each of the family candidates. Even with this extension of adjustable parameters, we obtained only $30$\% of converging orbits at the $7.6$~Myr epoch in the past. A more detailed analysis of the Koronis2 family's age is therefore needed and deferred to future work.

For now, we assume that the preselected group of candidate asteroids represents genuine members of the Koronis2 family. The left panels of Fig.~\ref{fig10} show these candidates projected onto 2D planes defined by proper semimajor axis versus eccentricity and proper semimajor axis versus sine of inclination, while the right panel displays their cumulative size frequency distribution. The blue and red lines represent variants that include or exclude asteroids (7719), (33143), and (40510), whose membership in the Koronis2 and Karin families remains ambiguous.

Similar to the case of the Karin family, the size distribution of the Koronis2 family in the 0.8 to 1.5 km range can be fit with a power law, $N(>D)\propto D^{-\alpha}$, where $\alpha=3.95\pm 0.01$. This value is only slightly shallower than that reported by \citet{broz2024}. If the slope exponent was steeper, the Koronis2 population could surpass that of the Karin family at sizes below a few hundred meters. Interestingly, this hypothesis can be directly tested with the forthcoming data from the Vera Rubin Observatory and NEO Surveyor.

\subsection{Extrapolation of Karin and Koronis2 observed populations to smaller sizes}\label{sfdkk}

The cumulative SFD of multikilometer fragments in the Karin and Koronis2 families was found to follow a power-law $N(>D)\propto D^{-\alpha}$, with $\alpha>3$ (Fig.~\ref{fig10}). This trend cannot persist down to very small fragment sizes, as the total mass of the fragment population would quickly exceed the mass that could be plausibly be ejected from the parent body. This suggests that there must be at least one change in the slope exponent $\alpha$ at some sub-kilometer size. In fact, it is more likely that multiple such changes occur or that the power-law approximation is only locally valid and $\alpha$ varies as a more complex function of size $D$. 

Here we briefly explore extrapolations of the observed SFDs in the identified Karin and Koronis2 families down to the meter-size range.  We consider the simplest scenarios, namely broken power-law distributions with one or two transitions to different values of $\alpha$. 

Let us approximate the population of the largest family members with $N(>D)=(D_1/D)^\alpha$. For Karin and Koronis2 families, we find $D_1\simeq 7.93$~km and $D_1\simeq 4.4$~km, with the slope exponents having the values given above. Formally extending the SFD beyond $D_1$ to infinity, the collective volume of fragments with a size larger than $D$ is $V(>D)=V_1\,\phi_V(D;D_1,\alpha)$ with
\begin{equation}
 \phi_V\left(D;D_1,\alpha\right) = \frac{\alpha}{\alpha-3}\,\left(\frac{D_1}{D}\right)^{\alpha-3}\; , \label{vol1}
\end{equation}
and the volume-equivalent size of spherical body they represent is $D_{\rm eff} = D_1\,\phi_V^{1/3}$. Here we see that $D$ cannot follow this trend to small sizes or both $V(>D)$ and $D_{\rm eff}$ would formally diverge in the limit $D\rightarrow 0$.

In order to keep the total volume of the fragments finite, we assume that the SFD becomes shallower below a certain break point $D_{\rm bp}$ with a power-law slope $\beta<3$. It can be shown that the number of fragments with a size larger than $D<D_{\rm bp}$ is $N(>D)=N(>D_{\rm bp})\,\psi_N(D;D_{\rm bp},\alpha,\beta)$ and their volume is $V(>D)=V(>D_{\rm bp})\,\psi_V(D;D_{\rm bp},\alpha,\beta)$, where
\begin{eqnarray}
 \psi_N\left(D;D_{\rm bp},\alpha,\beta\right) &=& 1+\frac{\alpha}{\beta}\left[\left(\frac{D_{\rm bp}}{D}\right)^\beta-1\right] , \label{pop2} \\
  \psi_V\left(D;D_{\rm bp},\alpha,\beta\right) &=& 1+\frac{\alpha-3}{3-\beta}\left[1-\left(\frac{D}{D_{\rm bp}}\right)^{3-\beta}\right]. \label{vol2}
\end{eqnarray}
The model may be easily generalized to an arbitrary number of breakpoints that separate different power-law segments of the SFD.

We now examine the hypothesis proposed in \citet{broz2024}, which suggests a link between the fragment populations of the Koronis2 (and Karin) families and the source of H-chondrite meteorites. This source was calibrated in that study to consist of approximately $(1-2)\times 10^{12}$ meter-sized main belt fragments. Our objective is to bridge this reference point with the observed kilometer-sized populations in the Karin and Koronis2 families using one or two breakpoint power-law approximations.

The constraints to be met are as follows. The modeling work of \citet{nes2006} estimates the parent body of the Karin family had an original size of approximately 33 km. Given the current size of (832) Karin, about 16.3 km, this leaves an available fragment population volume equivalent to a sphere with an effective diameter of $D_{\rm eff}\simeq 31.6$~km.  

\begin{figure}[t!]
 \begin{center}
  \includegraphics[width=0.46\textwidth]{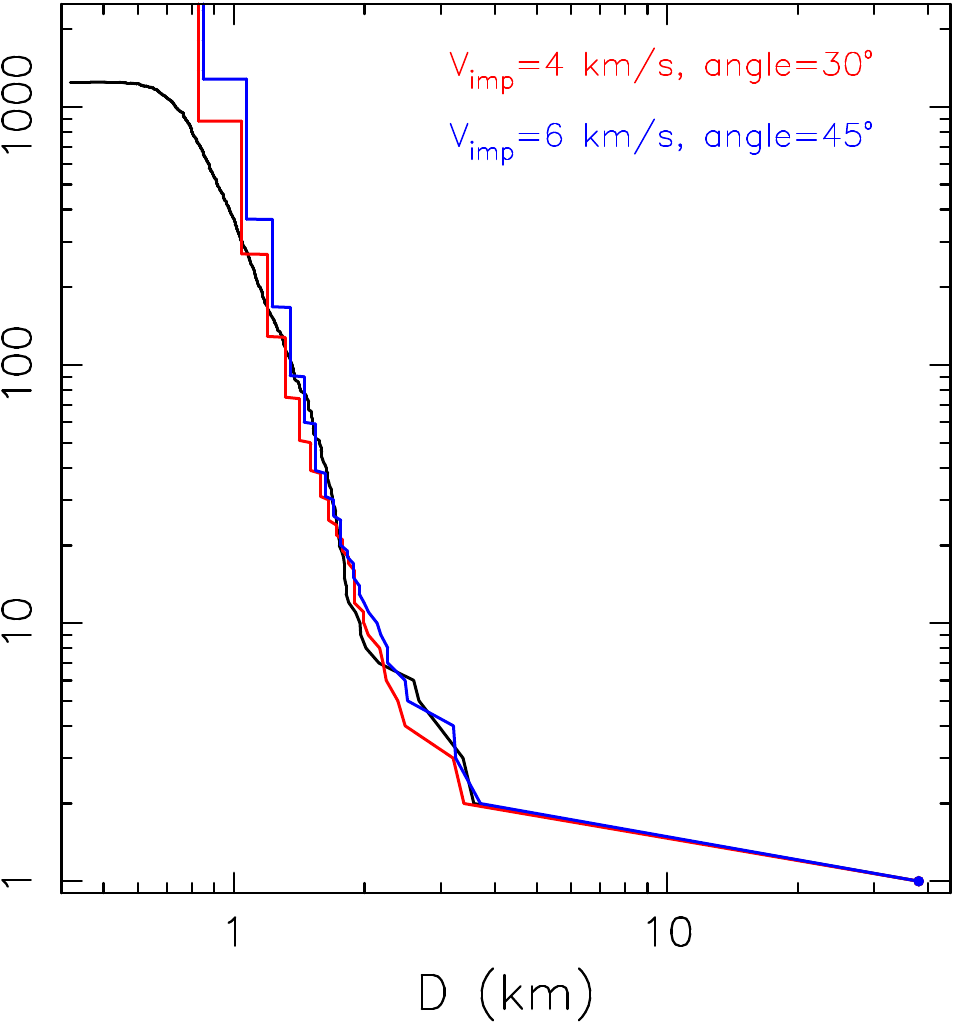}
 \end{center}
 \caption{Comparison of the Koronis2 family population (black) with two results of basaltic asteroid breakup database obtained in \citet{durda2007}. In both cases, the size of the largest fragment was scaled to $38$~km (assumed size of 158~Koronis) and the projectile size was $2.6$~km. The small difference of predicted fragment populations is due to different impact angle, $30^\circ$ (red) vs $45^\circ$ (blue), and impact speed, $4$ (red) vs $6$ km~s$^{-1}$ (blue). The equivalent volume of the fragment populations, starting from the second largest body, is $19$ and $23$~km.}
 \label{fig11}
\end{figure}
\begin{figure*}[t!]
 \begin{center}
  \includegraphics[width=0.9\textwidth]{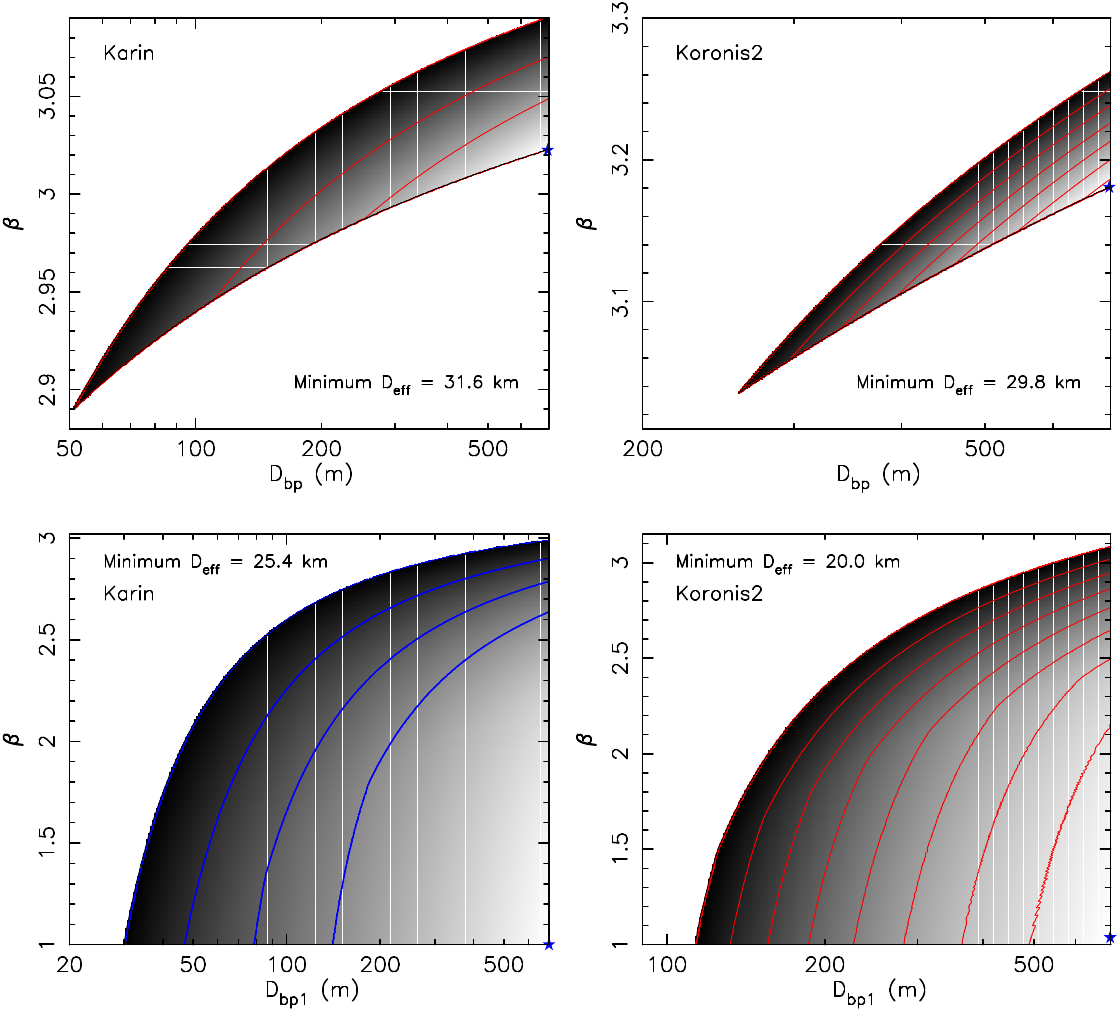}
 \end{center}
  \caption{Upper panels: Karin and Koronis2 extrapolated SFDs to meter size using a broken power-law with a break-point at $D_{\rm bp}$ (abscissa) and slope index $\beta$ for $D<D_{\rm bp}$ (ordinate). The shaded area delimits possible solutions with the following constraints: (i) $N(>1\,{\rm m})\geq 10^{11}$ and $D_{\rm eff}\leq 33$~km for the Karin family, (ii) and $N(>1\,{\rm m})\geq 2\times10^{11}$ and $D_{\rm eff}\leq 33$~km for the Koronis2 family. The minimum possible solutions for $D_{\rm eff}$ are located by blue star and the values given by label. The red isolines in shaded region are for $D_{\rm eff}=33$~km (at the upper zone of the admissible solution) decremented by $0.5$~km. In both cases the simple model with only one break point produces a tension by requiring too large minimum $D_{\rm eff}$ value.
  Lower panels: As above, but now for a piecewise power law SFD model with two break-points: the abscissa shows the value of the first breakpoint $D_{\rm bp1}$ the ordinate is the slope $\beta$ for $D_{\rm bp2}<D<D_{\rm bp1}$. The shaded area delimits possible solutions with the following constraints: (i) $N(>1\,{\rm m})\geq 10^{11}$ and $D_{\rm eff}\leq 31$~km for the Karin family, (ii) $N(>1\,{\rm m})\geq 2\times10^{11}$ and $D_{\rm eff}\leq 28$~km for the Koronis2 family (in each of the cases location of the second break point $D_{\rm bp2}$ and the slope index $\gamma$ for $D_{\rm bp2}$ are chosen to satisfy the constraints). The minimum possible solutions for $D_{\rm eff}$ are located by blue star and the values given by label. The blue isolines in shaded region for Karin are for $D_{\rm eff}=30, 29$ and $28$~km solutions. The red isolines in shaded region for Koronis2 are for $D_{\rm eff}=28$~km (at the upper zone of the admissible solution) decremented by $1$~km.}
 \label{fig12}
\end{figure*}
\begin{figure}[t!]
 \begin{center}
  \includegraphics[width=0.46\textwidth]{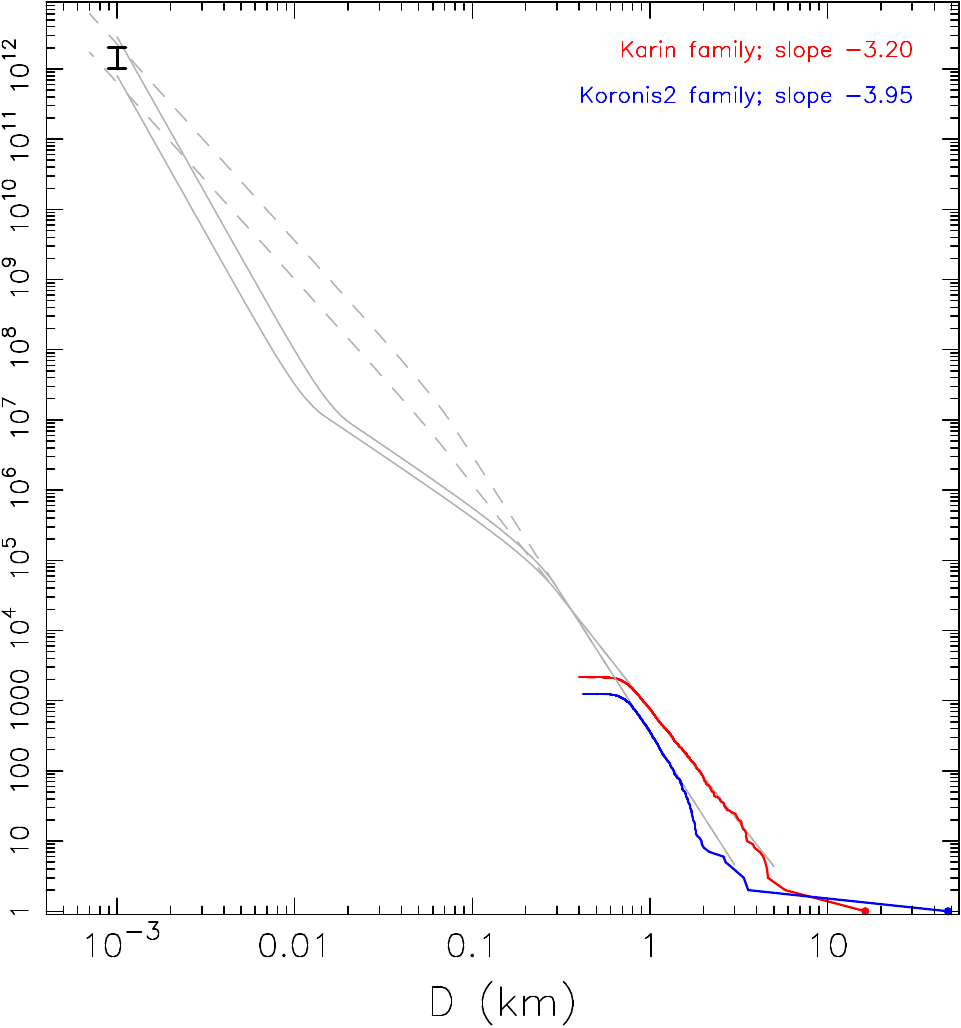}
 \end{center}
  \caption{Linking the size-frequency distributions (SFDs) of the identified Koronis2 (blue) and Karin (red) families is essential to establish them as plausible sources of the IRAS $2.1^\circ$ dust band (off-scale for size here) and H-chondrite meteorites. The slope exponents included in the labels correspond to the observed SFDs of the families down to $\simeq 0.6-0.7$~km. The solid gray lines represent extrapolations of the Koronis2 and Karin family SFDs, using two break points between power-law segments that satisfy the following criteria: (i) they reach the meter-sized population necessary to account for the observed flux of H-chondrite meteorites (black interval), and (ii) they maintain a reasonably small total volume, equivalent to a spherical body with a diameter of 24 km for the Koronis2 family and 27 km for the Karin family, respectively. In contrast, the previously suggested extrapolations with a single break point (illustrated by the dashed lines) consistently exceed the total available fragment volume (estimated here as 37 km for Koronis2 and 32 km for Karin). This analysis indicates that a wavy pattern in the SFD, spanning from meter-sized objects up to approximately $0.5$~km, is necessary to reconcile the observations. On top of that, further breaks in the distribution would be needed below $1$~m to link its value to the amount of dust needed in the dust band.}
 \label{fig13}
\end{figure}

However, the Karin family is also associated with the IRAS $\beta$ dust band \citep[see][]{nes2003,nes2006bands}, so the fragment SFD must extend down to micrometer sizes. According to \citet{nes2006bands}, the volume of particles within the $\beta$ dust band, spanning sizes from microns to centimeters, is equivalent to a sphere with a diameter of roughly 11 km. Even when accounting for the collisional evolution of these particles since the formation of the Karin family, as well as the uncalibrated segment of particles between a centimeter and a meter in size, we estimate that the total equivalent volume of fragments should not be greater than $\simeq 28-30$~km.

Less information is available for the Koronis2 family, partly due to the uncertainty in the size, $D_{158}$ of (158)~Koronis itself. For now, we rely on the numerical results from \citet{durda2007}, who conducted an extensive suite of smoothed particle hydrodynamic (SPH) simulations involving impacts between projectiles and basaltic targets of $D = 100$~km. Here \citet{durda2007} varied projectile sizes, impact geometries, and collsion velocities to generate a database of synthetic fragment populations that could be compared to observed asteroid families. Crucially, the simulated fragment populations conserved the total mass available in a given collision. By scaling these results to match the SFD of the largest fragments, presumably observationally complete, the simulations allow the total mass or equivalent volume of fragments to be calibrated.

Figure~\ref{fig11} presents two examples of synthetic fragment populations that closely match the Koronis2 SFD for fragments $\geq 1.1$~km in size. While the resolution of the \citet{durda2007} simulations does not permit precise comparisons for fragments smaller than $1$~km, the zero-order estimate of the total fragment volume remains valid (though we admit this conclusion needs to verified by the future SPH simulations with much larger resolution).

The two simulations shown in Fig.~\ref{fig11} correspond to equivalent fragment volumes associated with spheres of $19$ and $23$~km in diameter, starting from the second-largest object in the family. This estimate assumes a size of $38$~km for (158) Koronis. We verified that assuming a larger size for (158) Koronis results in a correspondingly smaller total volume for Koronis2 fragments.

We first tested a single-breakpoint solution for the SFDs of the Karin and Koronis2 families in the range between meter and kilometer sizes. The results are shown in the upper two panels of Fig.~\ref{fig12}, where the axes represent combinations of $(D_{\rm bp},\beta)$ values that satisfy the required constraints. 

We find that our solutions do a poor job of matching constraints, especially for the Koronis2 family: (i) $\beta>3$ is consistently required, which shifts the problem of an overly massive SFD tail to sizes smaller than a meter, and (ii) the minimum possible $D_{\rm eff}$ is $\simeq 29.8$~km. Given our estimate of the total fragment population volume for Koronis2, equivalent to a sphere with a diameter of 19 to 23 km (starting from the second-largest body), the single-breakpoint power-law model is not viable.

This issue is mitigated under the assumption of a broken power-law model with two breakpoints within the meter-to-kilometer size range, but at the cost of introducing more parameters and relying on less observational evidence. The lower two panels of Fig.~\ref{fig12} display possible combinations of $(D_{\rm bp1},\beta)$ values for the first segment, which must then be complemented by fine-tuned values of $(D_{\rm bp2},\gamma)$ for smaller fragment sizes. Under this two-breakpoint model, many solutions with $D_{\rm eff}<23$~km become viable. Figure~\ref{fig13} presents exemplary test cases for both the single- and two-breakpoint solutions.

The key takeaway from our analysis is as follows. While high yields of small fragments from recent breakups may be possible, the limited amount of source mass produced by a collision event imposes a significant constraint. Accordingly, extrapolations from the largest family members often result in an excessively large cumulative SFD of small fragments. If so, these solutions must be rejected. 

Ideally, introducing 'wiggles' into the fragment SFD could help address this problem. The issue is that such variations are not yet constrained by direct observational data. By incorporating these variations, one risks introducing a degree of arbitrariness into the results that could be problematic. For instance, it is plausible that all sorts of recent asteroid breakups could produce similar quantities of small fragments if their ejecta SFDs were allowed to have flexible shapes.

\section{Discussion and conclusions} \label{concl}
The new catalog of asteroid proper elements allowed us to substantially improve our knowledge of the Karin family. The principal results are as follows.
\begin{itemize}
\item The up-to-date family contains $2,161$ members, nearly five times more than the previously published census.

\item The cumulative SFD of the Karin family between $\simeq 0.8$ and $\simeq 3$~km is well approximated by a power-law with a slope $3.20\pm 0.01$.

\item The excellent convergence of proper nodal and perihelion longitudes of Karin members, with a dispersion of $\simeq 1^\circ$, can be achieved using reasonably constrained values of the semimajor axis drift rate ($da/dt$) caused by Yarkovsky thermal accelerations. For sub-kilometer Karin members, the $da/dt$ values exhibit a bimodal distribution, with roughly symmetric peaks corresponding to about half of the maximum expected value for their size and a minimum near zero drift. This result provides strong evidence that the YORP effect has tilted the rotation pole directions of these asteroids away from the ecliptic plane. The maxima of $da/dt$ distribution are located near effective obliquities of approximately $\simeq 60^\circ$ and $\simeq 120^\circ$, but due to use of a simplified constant drift rate $da/dt$ in our method, the exact obliquities of individual Karin members might be somewhat spread around these values.
  
\item The parameters suitable for achieving secular angle convergence enable us to estimate the initial configuration of the Karin family in the space of proper elements $(a,e,\sin I)$. The results closely mimic data obtained from numerical simulations of the parent asteroid's breakup that originally formed the Karin family. Considering the much larger number of Karin family members known today, it would be worthwhile to revisit and refine the breakup simulations in the near future. 

\item We proposed a new metric function for orbital convergence, specifically designed for young asteroid families, which offers several advantages over the simple convergence of nodal and perihelion longitudes. This new approach eliminates singularities at zero eccentricity or inclination and directly relates to the properties of the initial ejection velocity field responsible for fragment dispersion. Additionally, we implemented an online determination of mean orbital elements and tested this framework on the Karin family. This analysis yielded an estimated age of $5.72\pm 0.09$~My (95\% C.L.) and expected ejection velocities of $\simeq(10-20)$ m~s$^{-1}$ for kilometer size fragments.
\end{itemize}

An intriguing puzzle persists regarding the population of large Karin members. Specifically, three asteroids, (7719), (33143), and (40510), exhibit convergence with (832) Karin at the formation epoch of the Karin family, but they also converge with (158) Koronis at epochs beyond 10 My, with the exception of (40510), which would require the Koronis2 family to have an age exceeding 15 My. Interestingly, determining their rotation poles could provide an effective means of distinguishing their family membership. For instance, if (7719) and (33143) belong to the Koronis2 family, and its age is $\simeq 13$ My, their pole obliquities would need to be $\le 40^\circ$. Conversely, if they are members of the Karin family, their rotation poles would likely align closer to the ecliptic plane. These predictions can be tested through an analysis of photometric observations.

Several upcoming small body surveys, such as the Vera Rubin Observatory and the NEO Surveyor mission, are poised to elevate our understanding of the Karin family to an entirely new level. For instance, the Vera Rubin Observatory is expected to complete the census of the main belt population down to $H\simeq 18.5$ \citep[e.g.,][]{lsst2025}. Assuming a geometric albedo of 0.23, this magnitude limit corresponds to an asteroid size of approximately 0.5 km. If the power-law slope of the SFD remains $\alpha\simeq 3.2$ all the way to this size limit, the observable population of the Karin family could increase to nearly 7,000 objects. Alternatively, the SFD of Karin family might start to bend shallower between $\sim(0.3-0.5)$~km and this feature could be detectable in the LSST data \citep[note that existence of such a breakpoint has recently been reported for very young families in the inner main belt, see][]{vok2024}.

Extending the analysis of the size dependence of mean $da/dt$ values required for convergence (as shown in Fig.~\ref{fig6}) to sizes down to $D\simeq 0.5$~km might yield valuable insights into the degree of collisional activity and the characteristic YORP timescale for sub-kilometer asteroids in the main belt. For example, \citet{bot2005} predicts a collisional lifetime of approximately 100 My for a 0.5 km asteroid. Statistically, this suggests that around $\simeq 6$\% of the smallest known Karin members should have undergone a catastrophic collision. Similarly, the calibration of mean YORP strengths by \citet{car2016} for multikilometer Karin members indicates that the YORP timescale for 0.5 km objects should be as short as $\simeq 15$~My. The combined effects of these two processes will largely determine the degree of polarization of $da/dt$ in Fig.~\ref{fig6} at such small sizes. 

Another approach to estimate the average strength of the YORP effect would be to determine the fraction of asteroid pairs among the Karin family members \citep[e.g.,][]{vok2008}. This is because such pairs are predominantly formed through YORP-driven rotational fission of their parent asteroids \citep[e.g.,][]{pra2010}. While we did not conduct a systematic search for asteroid pairs within the current Karin population in this study, we note one remarkable example: asteroids (82780) 2001 QF18 and (604365) 2015 PO96. These two objects appear to form a primary-secondary pair, and are estimated to be $\simeq 100$~ky old. The primary is $\simeq 2.6$~km in size, while the secondary has an approximate size of 1 km. Beyond this, no additional data on their physical properties, such as rotational periods, are currently available. Future observations from upcoming large-scale surveys, which will greatly expand the known Karin population, are expected to reveal many more such cases.

The accurate determination of the Karin family has allowed us to better isolate the candidate population of the partially overlapping Koronis2 family compared to previous studies. The kilometer-sized population of the Koronis2 family exhibits a very steep SFD, approximated by a power law with a cumulative slope of $3.95 \pm 0.01$. We attempted to determine the age of the Koronis2 family using the simple convergence of proper node and perihelion longitudes with (158)~Koronis. Unfortunately, the situation is more complex than in the case of the Karin family. While the age of $7.6$~My proposed by \citet{broz2024} is plausible, it is challenged by the apparent lack of convergence among some large candidate members.

In our analysis, this issue can be resolved if the Koronis2 family is assumed to be older than $7.6$~My. If so, the Koronis2 family may no longer be a good candidate to explain the large fraction of H chondrites with cosmic ray exposure ages near this age \citep[e.g.,][]{mg1992,gm1995}. Alternatively, the limitations of our methodology could mean that the Koronis2 age must be determined through direct numerical simulations. When this is revisited in the future, we recommend using the new convergence metrics outlined in Appendix~\ref{kamean}. 

With our updated membership for the Karin and Koronis2 families, which are presumed complete down to approximately $1$~km in size, we briefly revisited extrapolations to the meter-size population, particularly in the context of their relevance as meteorite sources. We find that the threshold levels established in \citet{broz2024} can be met, but achieving this feat is challenging, if not impossible, under the simplest assumption of a broken power-law with two segments of constant slope indices and a single breakpoint. Such models tend to systematically predict excessively large fragment populations. Although the issue can be mitigated by introducing additional power-law segments, this approach introduces arbitrary elements into the models, making the proposed solutions less compelling.

\acknowledgments
The authors thank Miroslav Bro\v{z} for helpful discussions and the referee, Bojan Novakovi{\'c},
for suggestions that made us improve the original version of this paper. The work of DV was
supported by the Czech Science Foundation through grant 25-16507S, DN's and WB's work was
supported by the NASA's Solar System Workings program through grant 80NSSC21K1829.
 
\bibliographystyle{aasjournal}

\appendix

\begin{figure*}[t!]
 \begin{center}
  \includegraphics[width=0.9\textwidth]{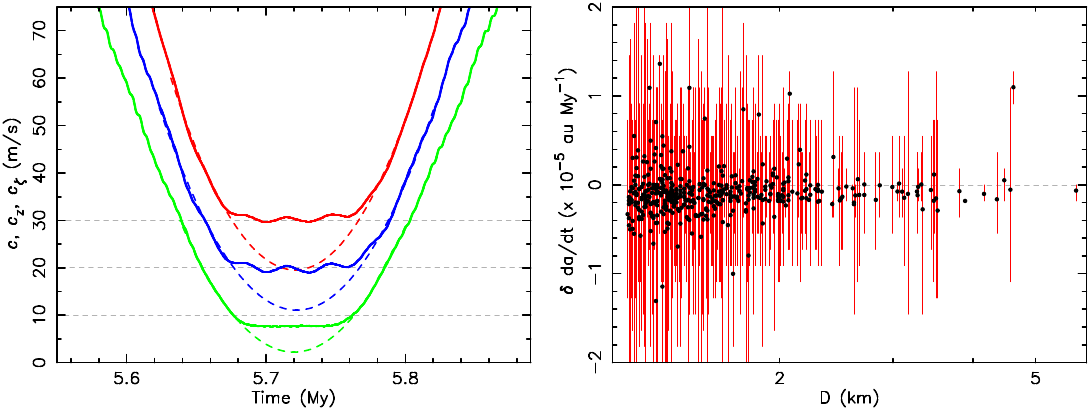}
 \end{center}
  \caption{Convergence of orbits of the $450$ largest Karin family members using mean orbital elements. The left panel shows quantitative indicators of the success expressed using metrics (\ref{tf1}) normalized to one converging orbit and given in velocity units: (i) ${\bar \cC}$ combining contribution from all initial velocity components (red curve), (ii) ${\bar \cC}_{\bf z}$ depending only the in-plane radial and transverse ejection velocity components (blue curve), and (iii) ${\bar \cC}_{\boldsymbol{\zeta}}$ depending only on the out-of-plane ejection velocity component (green curve). The horizontal dashed lines provide reference level of $10$, $20$ and $30$ m~s$^{-1}$. Perturbations by massive bodies in the main belt, Ceres, Pallas or Vesta (not included in the simulation), prevent minimum values at $T\simeq 5.72$~My decrease deeper along the dashed lines representing  a quadratic fit of the wings of the functions. The wavy pattern in ${\bar \cC}_{\bf z}$, reflected also on ${\bar \cC}$, is likely produced by perihelion precession of Karin's orbit (see the main text). The right panel shows slight adjustments of the semimajor axis drift-rates $\dot{a}$ for individual orbits compared to their a priori solutions from the convergence of the orbits using only proper node and perihelion values (shown in Figs.~\ref{fig3} and \ref{fig6}) as a function the asteroid size $D$. These corrections are very small, less than $10$\% of the apriori values. The black symbols are the mean, and red vertical bar is the range between minimum and maximum values, of $\dot{a}$-corrections for epochs $T$ between $5.69$ and $5.74$~My.}
 \label{figKamean}
\end{figure*}

\section{Convergence of large Karin members in mean orbital elements}\label{kamean}
In this Appendix, we explore the possibility to use mean, instead of proper, orbital elements for convergence studies of young asteroids families and illustrate our approach in the case of the Karin family.
While not new in this respect \citep[see, e.g.,][]{nov2012,nov2014}, our goal is to (i) remove singular behavior of the secular angles when eccentricity or inclination become very small, and (ii) achieve a stronger link to the initial configuration of the family than provided by simple convergence of the secular angles and to obtain higher precision in its reconstruction (Sec.~\ref{ka3}). These high ambitions require one to switch from an analytical to a numerical approach to the problem. Yet, results obtained in Sec.~\ref{karin} provide us with important initial information. This includes: (i) identification of Karin family members, and (ii) pre-constrained values of the optimum model parameters that point to the family origin at $T\simeq 5.72$~My (such as the $\dot{a}_j$ of each of the family asteroids). 
\smallskip

\noindent{\it Numerical integrator. }The numerical simulations were performed using the well tested integration package {\tt swift}%
\footnote{\citet{ld1994} and \url{http://www.boulder.swri.edu/~hal/swift.html}.}
that we complemented with two extensions. The first emulates the Yarkovsky effect using a simple along-track acceleration that results in the expected secular semimajor axis drift $\dot{a}$. This part has been used and tested repeatedly in a number of our previous studies of asteroid families \citep[e.g.,][]{vok2006}. The new feature is the on-line computation of the mean orbital elements.%
\footnote{See also an extensive software package by Mira Bro\v{z} available at GitHub, which also includes on-line digital filtering and many more options (\url{https://github.com/miroslavbroz/swift_rmvs3_fp_ye_yorp}).}
We adopted the AAAB sequence of low-frequency filters described and tested in \citet{quinn1991} and implemented them as an on-line tool into the swift package. We used a $6$~d timestep and $36$~d sampling frequency of the bottom A-filter. As suggested in \citet{quinn1991} we used a shift factor of $10$ before applying the next filter in the hierarchy. With that scheme applied, the signal with periods less than $\simeq 660$~yr is suppressed with a factor better than $10^{-4}$, while the secular variations of the orbital elements with periods larger than $\simeq 2$~kyr are effectively preserved. This is an excellent realization of the numerical transformation between the osculating and mean orbital elements. The initial heliocentric state vectors for all planets and asteroids were obtained from the JPL Horizons website and correspond to the initial epoch MJD 60800. They were given in the ecliptic system of J2000.0, but before the propagation we rotated them into the invariable (Laplacian) system. In order to represent backward orbital evolution we inverted orbital velocities and the semimajor axis drift-rates $\dot{a}$ for all asteroids. We used a priori values from the convergence solution in proper orbital elements (Sec.~\ref{karin}), but allowed slight adjustments within $\pm 2\times 10^{-5}$ au~My$^{-1}$. 
\smallskip

\noindent{\it Target function. }Since the mean orbital elements contain more detailed information about the orbital evolution than the proper elements used in the main text, in particular the mean eccentricities $e$ and inclinations $I$ are not constant, the target function (\ref{tf}) for the minimization algorithm to track the family origin should be replaced with an appropriately more complex form, now depending also on the behavior of $e$ and $I$. Equations (\ref{gau4}) and (\ref{gau5}), characterizing the change in the orbital elements at the origin and its direct relation to the velocity field, provide the basis. Since the target function must be positive definite, we use
\begin{equation}
 \cC^2\left(T\right) = \cC^2_{\bf z}+\cC^2_{\boldsymbol{\zeta}} =\sum_{j=1}^N\left[\Delta{\bf z}_j\,\Delta{\bar{\bf z}}_j+\Delta\boldsymbol{\zeta}_j\,\Delta\bar{\boldsymbol{\zeta}}_j\right]
 \; , \label{tf1}
\end{equation}
where $\Delta{\bf z}=\Delta k+\imath\,\Delta h$ and $\Delta\boldsymbol{\zeta}=\Delta q+\imath\,\Delta p$ are the numerically inferred differences of the mean orbital elements with respect to (832)~Karin (the overbar means a complex conjugate quantity). If the correspondence to the ejection velocities would hold the same as for the osculating orbital elements, the target function (\ref{tf1}) would have the following interpretation in the velocity field components: $\cC^2=\sum_{j=1}^N\left[v_r^2+4v_t^2+v_z^2\right]_j$ valid to the zero order in orbital eccentricity $e$.%
\footnote{Developing $\cC^2$ to the first power in orbital eccentricity, and dropping terms depending on orbital true anomaly (information not accessible by the mean orbital elements), we obtain $\cC^2=\sum_{j=1}^N \left[v_r^2+v_r v_t\, h_\star+ 4v_t^2\left(1-\frac{1}{2} \,k_\star\right)+v_z^2\right]_j$, where $k_\star$ and $h_\star$ are the non-singular variables of the reference orbit (such as 832~Karin).}
Similarly, the $\cC^2_{\bf  z}=\sum_{j=1}^N\left[v_r^2+4v_t^2\right]$ and $\cC^2_{\boldsymbol{\zeta}}=\sum_{j=1}^N v_z^2$ parts would characterize contributions of the in-plane and out-of-plane velocity components. Analyzing the two parts separately offers the possibility --at least in principle-- to probe various coarse hypotheses about the ejection velocity field. For example, consider that fragments were ejected with a characteristic velocity $v$ and directed uniformly in conus with an opening angle $\theta_\star$ in the out-of-plane direction. Then, $\cC^2_{\bf z}=\frac{5}{3}\,v^2\left(1-\frac{1}{2}\,\phi_\star\right)$ and $\cC^2_{\boldsymbol{\zeta}}=\frac{1}{2}\,v^2 \left(1+\phi_\star\right)$ with $\phi_\star=\cos\theta_\star\left(1+\cos\theta_\star\right)$. If the conus is directed along the transverse direction in the orbital plane, the weight changes in favor of $\cC^2_{\bf z}$, and one has
$\cC^2_{\bf z}=\frac{5}{3}\,v^2\left(1+\frac{7}{25}\,\phi_\star\right)$ and $\cC^2_{\boldsymbol{\zeta}}=\frac{1}{3}\,v^2 \left(1-\frac{1}{2}\phi_\star\right)$.
An isotropic ejection field, $\theta_\star=\pi$, yields in both cases $\cC^2_{\bf z} = 5\,\cC^2_{\boldsymbol{\zeta}}$, and the same result is also obtained for the hemispheric ejection fields, $\theta_\star =\pi/2$. This degeneracy is caused by only quadratic velocity-component contribution in $\cC^2$. For sake of reporting our results in units of velocity, we also define ${\bar \cC}=V_0\sqrt{\cC^2/N}$,  ${\bar \cC}_{\bf z}=V_0\sqrt{\cC^2_{\bf z}/N}$ and ${\bar \cC}_{\boldsymbol{\zeta}}=V_0\sqrt{\cC^2_{\boldsymbol{\zeta}}/N}$ with $V_0=17600$ m~s$^{-1}$.
\begin{figure*}[t!]
 \begin{center}
  \includegraphics[width=0.9\textwidth]{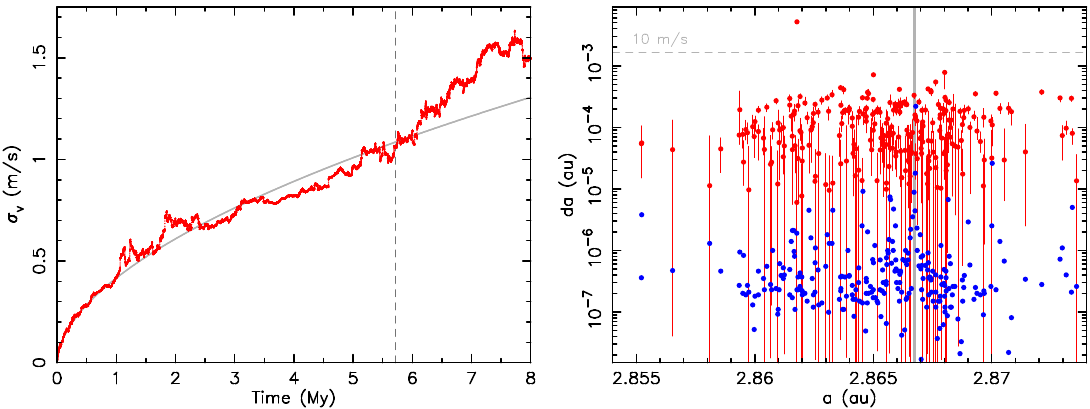}
 \end{center}
  \caption{The effect of massive bodies in the asteroid belt -- (1)~Ceres, (2)~Pallas and (4)~Vesta (CPV) -- on mean semimajor axis of members in the Karin family. Left panel: Growth of the semimajor axis dispersion expressed by $\sigma_a^2(T)=\langle [a(T)-a(0)]^2\rangle$ with time $T$ translated to characteristic velocity perturbation using $\sigma_v=V_0\,\sigma_a/a_0$ ($V_0=17600$ m~s$^{-1}$ and $a_0=2.865$~au). An approximate power-law fit $\sigma_v\propto T^{0.66}$ is shown by the gray line. Orbits that underwent only distant encounters with CPV contribute to this statistics, as those suffering deep encounters do not obey statistical properties of a quasi-random walk. The vertical dashed line marks the $5.72$~My age of the Karin family.
  Right panel: The change in the mean semimajor axis $da(T)=|a(T)-a(0)|$ for $T$ in between $5.5$ and $6$~My for orbits of $252$ largest members in the Karin family: red symbols are the median and lines show the range of minimum to maximum value \citep[the vertical gray line denotes location of a 3 body mean motion resonance $4\lambda-13\lambda_J+8\lambda_S+\varpi$ with Jupiter and Saturn, see][]{gal2014}. The blue symbols are formal standard deviations of the proper elements from \citet{nes2024}. The dashed horizontal line corresponds to an equivalent velocity kick of $10$ m~s$^{-1}$. Orbits contributing above this threshold ($1$ out of $252$ over $6$~My) imply deep encounters with CPV.}
 \label{figcpv}
\end{figure*}
\smallskip

\noindent{\it Preliminary results. }Figure~\ref{figKamean} shows the results for $450$ largest members in the Karin family. The left panel expresses their convergence using the new metrics ${\bar \cC}$ in (\ref{tf1}), and the related parts ${\bar \cC}_{\bf z}$ and ${\bar \cC}_{\boldsymbol{\zeta}}$ depending on the in-plane and out-of-plane ejection velocity components. They can be directly compared with $\cC^\prime$ in Fig.~\ref{fig2} since at each past epoch $T$ we adjusted individual drift rate values $\dot{a}_j$ to achieve the optimum convergence level. The minima of all metric functions are now appreciably narrower in $T$, and the expression of a higher-fidelity convergence procedure. However, at their minima, none of the ${\bar \cC}$-functions is matched by a quadratic function in $T$ as expected (see the fits shown by the dashed lines). We believe that this behavior has to do with perturbation by massive objects in the main belt, dominantly Ceres and Pallas, which are not included in our simulation, yet they affected the present-day orbits of Karin members. In the next section, we briefly examine their expected level and conclude that it may easily explain the few meter per second deficiency in minima seen in Fig.~\ref{figKamean}. If we were to consider the ideal quadratic profile from the ${\bar \cC}_{\boldsymbol{\zeta}}$ component and $\simeq 5$ m~s$^{-1}$ noise (at the level of the minimum deficiency), the formal Gaussian statistics would set the $95$\% C.L. for the Karin family age solution at $5.72\pm 0.09$~My. The wavy pattern at the minima of the ${\bar \cC}$ functions is an interesting characteristic, most likely an expression of first order in $e$ terms (see footnote~3). Analyzing its amplitude may provide further information about the statistical distribution of the radial and transverse ejection velocity fields. The right panel of Fig.~\ref{figKamean} shows the range of adjustments $\dot{a}$ required to achieve the best convergence in mean orbital elements compared to those following from proper orbital elements (Sec.~\ref{karin}). These corrections are very small and only weakly dependent on the size $D$, indicating that the solution followed from the convergence of the proper secular angles is already a very good approximation.

The convergence approach in the space of mean orbital elements briefly outlined in this Appendix seems to us a promising novel tool that directly reaches quantitative aspects of the initial ejection field of fragments in the family. Our goal here was to introduce it and use it to justify results in the main body of the paper, but not to elaborate on its details. We postpone this work to future publications.

\subsection{Perturbations from massive bodies in the main belt}\label{seccpv}
In order to probe the role of the gravitational perturbations by most massive bodies in the main belt in the Karin and Koronis2 orbital zone, we conducted the following experiment. We selected the largest $252$ asteroids in the Karin family, roughly uniformly distributed in proper $a$ and $e$ values. We set their semimajor axis drift rates to zero and propagated their orbit for $10$~My backward in time. If only planets were included, the mean semimajor axes $a$ of the family members should have been constant (except for resonant effects, which are tiny in this orbital region). In addition to planets, we also included three perturbers embedded in the main belt, namely (1)~Ceres, (2)~Pallas and (4)~Vesta (CPV), accounting for more than $50$\% of its mass. While less massive than planets, the nature of their gravitational influence on members in the Karin family is significant and consists of two distinct processes: (i) a cumulative effect of distant encounters resulting in quasi-random walk in the orbital space, and (ii) stronger, but less frequent, perturbations due to deep close encounters. We illustrate their effects on the evolution of the mean semimajor axes of our sample of Karin members. To that end, we determine the change in the mean semimajor axis $\delta a_j(T)=a_j(T)-a_j(0)$ for each asteroid orbit and past epoch $T$, and from these values we compute the characteristic dispersion $\sigma^2_a(T)=\sum_j \delta a_j^2/N$ (which can also be expressed in terms of velocity $\sigma_v=V_0\, \sigma_a/a_0$, where $V_0=17,600$ m~s$^{-1}$ and $a_0=2.865$~au). The sum goes over $j=1,\dots,N$, where $N$ is the subsample of bodies undergoing only distant encounters (process --i-- above). We monitor the effects of deep encounters separately, identifying them with $|\delta a_j|>2\times 10^{-3}$~au.

Figure~\ref{figcpv} summarizes the results of our simulation. The left panel shows growth of the dispersion $\sigma_v(T)$ in the sample of orbits avoiding deep encounters with CPV. As already found in 
\citet{nes2002f} \cite[see also][]{car2003}, $\sigma_v(T)\propto T^\beta$ with $\beta\simeq 0.66$, slightly larger than $0.5$ expected for a random-walk process. This is likely because of the correlated or repeated encounters with CPV during favorable orbital alignments. The right panel shows the accumulated values $\delta a_j(T)$ for $T\in (5.5,6)$~My (red symbols). The majority of the data correspond to the population of orbits with only distant CPV encounter history, for which $\delta a_j \simeq (2-3)\times 10^{-4}$~au is appreciably greater than the formal uncertainty of the proper $a$ values determined in a dynamical model with only planetary effects included (blue symbols). We found that one orbit of the $252$ sample had a deep encounter with CPV within the $6$~My time interval. This case, with $\delta a > 2\times 10^{-3}$~au, is distinct from the other orbits. 

The takeaway message from our simulation is as follows. Perturbations from CPV, and other massive asteroids in the main belt, impose a noise in the orbital convergence efforts of young families. In the case of the Karin family, it is of the order of meters per second in each of the ejection velocity components, explaining the difference between the ideal minima and observed "stalled" minima of the ${\bar \cC}$-metric functions shown on the left panel of Fig.~\ref{figKamean}. Additionally, a few percent of Karin family objects might have suffered significant orbital change due to deep encounters with CPV that would even prevent their identification as its members from today's orbits.

\end{document}